\documentclass[twocolumn]{aastex631}
\usepackage[utf8]{inputenc}

\usepackage{lineno}

\begin{document}

\title{The Morphology of Dwarf Galaxies Hosting Variable Active Galactic Nuclei}
\author{Erin Kimbro }
\author{Vivienne Baldassare}
\author{Guy Worthey}
\affiliation{Department of Physics and Astronomy, Washington State University, Pullman, WA, 99163}
\author{Marla Geha}
\affiliation{Yale University, New Haven, CT 06511}
\author{Jenny Greene}
\affiliation{Princeton University, Princeton, NJ 08544}

\begin{abstract}
   We analyze Hubble Space Telescope (HST) optical imaging of eight low-mass galaxies hosting active galactic nuclei (AGN) identified via their photometric variability in \cite{baldassare_search_2020}. 
   We use GALFIT to model the 2D galaxy light profiles, and find a diversity of morphologies.
   The galaxies with regular morphologies are best fit with pseudo-bulges and disks, rather than classical bulges. We estimate black hole masses using scaling relations and find black hole masses of 10$^{3.7-6.6}$ M$_\odot$. We compare this sample to dwarf galaxies with AGN selected via optical spectroscopy. On average, the variable host galaxies have lower mass black holes. We analyze the brightest point source in each galaxy and find their properties are not entirely consistent with star clusters, indicating that they are likely AGN. These point sources are found to have lower luminosities than spectroscopically selected dwarf AGN, but brighter than the point sources in dwarf galaxies not identified as AGN. Our detailed imaging analysis shows that variability selection has the potential to find lower mass black holes and lower luminosity AGN than optical spectroscopy. These active dwarfs may have been missed by spectroscopic searches due to star formation dilution or low gas content. 
 
\end{abstract}

\section{Introduction}



Super massive black holes (SMBHs) are known to exist at the center of every massive galaxy \citep{magorrian_demography_1998}. The exact formation mechanism of SMBHs is not known, however the occupation fraction of intermediate mass black holes (IMBHs) ($M_{BH} < 10^6$) can provide observational constraints for different theories of MBH formation \citep{greene_low-mass_2012}. One method to search for IMBHs is to look for signatures of active galactic nuclei (AGN) in nearby dwarf galaxies. 

We are finding more evidence for active dwarf galaxies in systematic searches for AGN \citep{baldassare_search_2020, reines_dwarf_2013, sartori_search_2015}. However, their effect on the host galaxy and occupation fraction in these systems are not well constrained observationally. Additionally, the tests used to select AGN in more massive galaxies are not as successful in dwarfs with only moderate overlap between different selection methods \citep{wasleske_active_2024}. Understanding the hosts of these low mass systems can both provide insights to how black holes affect galactic environments and the physical processes in IMBHs. 


The search for IMBHs in dwarf galaxies ($M_{\ast}\lesssim3\times10^{9}\;\rm{M_{\odot}}$) can employ a variety of techniques, each with their own biases. The first large sample of AGN in dwarf galaxies was found by \cite{reines_dwarf_2013} using the BPT diagram. However, the BPT diagram may not be well suited to finding AGN in dwarf galaxies \citep{cann_limitations_2019, trump_biases_2015}. Calibrated on bright quasars to separate out star forming galaxies from AGN activity, it is biased against low-luminosity AGN and low-metallicity host galaxies \citep{groves_emission-line_2006}. Variability studies are a promising avenue to detect the presence of an AGN in dwarf galaxies that is less dependent on star formation and/or metallicity \citep{baldassare_identifying_2018, baldassare_search_2020, burke_dwarf_2022, ward_variability-selected_2022}. 

With the increase in the number of dwarf galaxies known to host AGN, it is now possible to explore the impact of galaxy structure and environment on the presence of an AGN (as well as the impact of the AGN on the host galaxy). Simulations have shown that AGN feedback maybe necessary to quench star formation in dwarfs \citep{koudmani_two_2022, arjona-galvez_role_2024}, though more observational constraints are needed, and 
some evidence of positive AGN feedback in dwarfs has also been found \citep{schutte_black-hole-triggered_2022}. 
Galaxy properties such as velocity dispersion, stellar mass, and bulge mass have also been linked to BH mass indicating that the evolution of the BH and the host galaxy are linked. 
However, at the lower mass end, it is unclear whether black hole mass continues to scale with galaxy properties \citep{jiang_black_2011}, and what host properties correlate with AGN activity in low mass galaxies. 

Several studies have explored the morphologies of dwarf galaxies with optical spectroscopic AGN signatures. \cite{greene_black_2008} studied the structure of IMBHs with broad emission lines and found that it was not necessary for the galaxy to have a classical bulge in order to host a MBH and their sample had a broad range of morphological types. This sample was unconstrained by stellar mass. They find the $M_{BH} - L_{\rm bulge}$ relationship breaks down as black hole mass decreases when compared to $M_{BH} - L_{\rm bulge}$ for more massive inactive galaxies. This is somewhat contrary to what is found in \cite{schutte_black_2019} which looks at the $M_{BH} - M_{bulge}$ relationship in dwarf galaxies hosting AGN. They find that their sample follows the same $M_{BH} - M_{bulge}$ scaling relationship of more massive galaxies and quiescent black holes.  \cite{jiang_host_2011} studied the structure of broadline selected AGN and found that most galaxies had pseudobulges and very few of them lived in classical bulges, indicating that IMBHs may have evolved secularly. \cite{kimbrell_diverse_2021} found similar results in their sample of BPT selected AGNs in dwarf galaxies, where most of the galaxies had pseudobulges, with a handful of exceptions. \cite{kimbrell_comparison_2023} explored this further by comparing their sample of AGN host galaxies to inactive dwarf galaxies. A large number of the inactive dwarf galaxies did not have any pseudobulge component, where it was very common for the active dwarf galaxies to have a pseudobulge, indicating that the two populations of dwarf galaxies have differing host galaxy morphology. Additionally, the central point sources in the active dwarfs were much more luminous than the inactive dwarfs. 

Most studies on galaxy structure and morphology of galaxies hosting IMBHs have focused on objects selected through optical spectroscopy or dynamical detections. 
How the morphology and environment of variability selected AGN in dwarf galaxies differ from objects selected with other techniques is important because the variability selected AGN may correspond to a different population of objects. 


This paper presents an analysis of the morphologies of a sample of dwarf galaxies with AGN discovered through their optical photometric variability.  
 We obtained Hubble Space Telescope (HST) imaging of eight dwarf galaxies hosting AGN initially identified in \cite{baldassare_search_2020}. All of these galaxies are selected as star-forming or composite galaxies by the BPT diagram. We present the first detailed study of the morphological properties of variability selected AGN and compare them to previous samples of dwarf galaxies with AGN selected through optical spectroscopy.

\section{Sample Properties and Observations}

The parent sample of the eight dwarf galaxies studied here were identified via their optical variability using the Palomar Transient Factory (PTF) \citep{baldassare_search_2020}. They identified 424 galaxies with AGN like variability covering a mass range of $10^7 - 10^{12} M_{\odot}$, { with stellar mass estimates taken from the NASA-Sloan Atlas (NSA) .} The eight galaxies chosen for follow up observations with \textit{HST} were chosen to have stellar masses less than $5\times10^{9} M_{\odot}$, $z<0.02$, and SDSS spectroscopy. We calculated the metallicity of these galaxies from SDSS spectra using pPXF \citep{cappellari_full_2023}. The galaxies are identified as star forming or composite by the BPT diagram or they are quenched. 

We observed eight dwarf galaxies with \textit{HST} in the mass range $M_* = 10^{7-9}$ with redshifts ranging from $z = 0.018 - 0.025$. Observations were taken using the Wide Field Camera 3 (WFC3) with the UVIS and IR channel from October 31st 2020 to April 21st 2021 (HST GO 16423). {The data presented can be obtained from the Mikulski Archive for Space Telescopes (MAST) and are available at dataset[doi: 10.17909/a3nk-5h54]{https://doi.org/10.17909/a3nk-5h54}.} Each galaxy was observed in the F110W (near-IR), F606W (optical), and F300X (near-UV) filter. The UVIS filter images have a pixel scale of $0.04"/$pixel and the IR image has a pixel scale of $0.13"/$pixel. Images were taken using a four-point dither patern and drizzled and reduced with the STScI astrodrizzle method. Total exposure times were roughly 200s, 500s, and 1400s in the F110W, F606W, and F300X filters, respectively. Galaxy properties from NASA-Sloan Atlas are listed in Table \ref{tab:prop}.


\begin{figure*}
    \centering
    \includegraphics[width=1.0\textwidth]{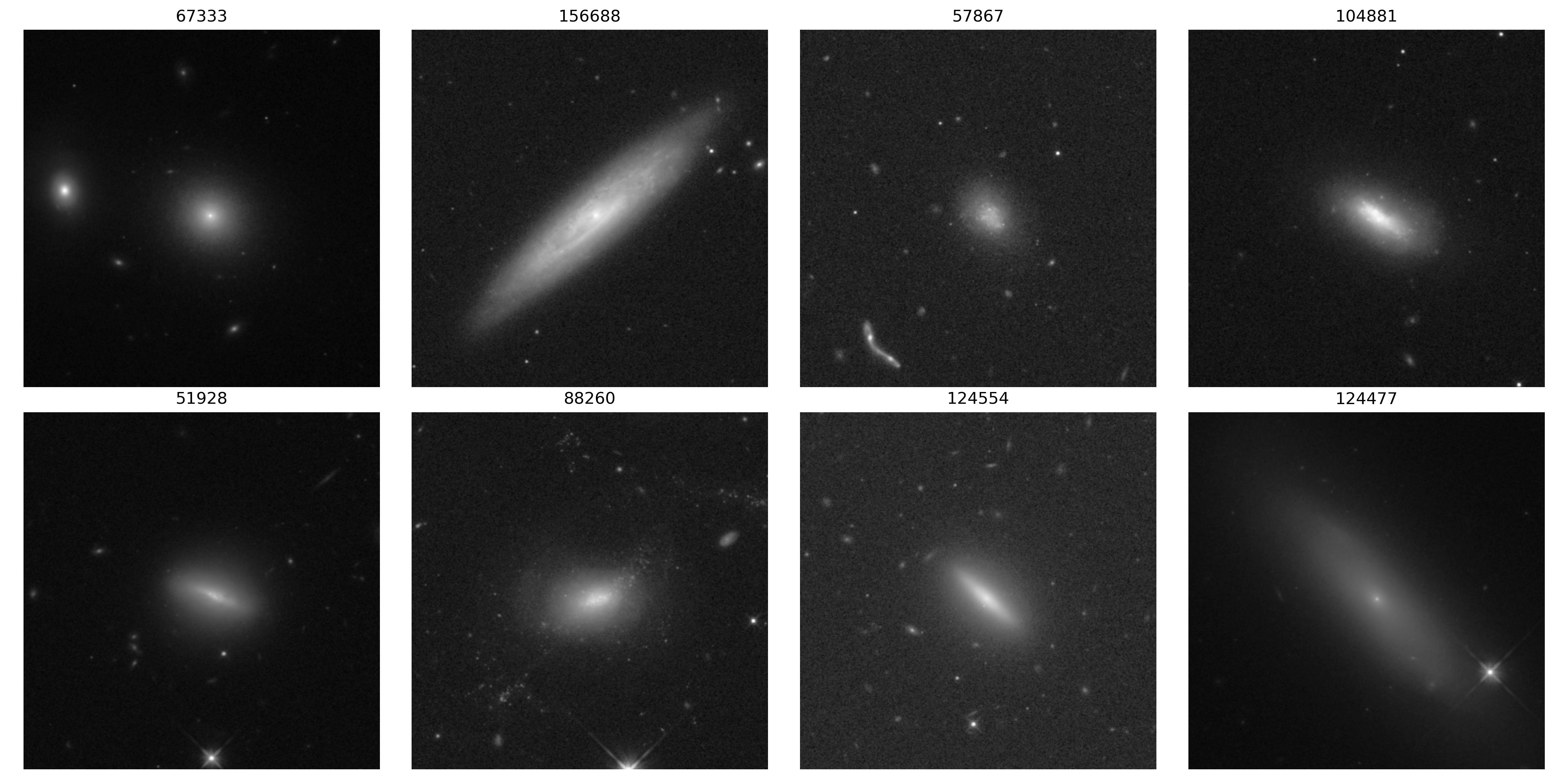}
    \caption{\textit{HST} F110W images for the eight dwarf galaxies in our sample. They are shown at the same scale at 0.13"/pixel and images are 52$''$ on each side.}
    \label{fig:110}
\end{figure*}

\begin{deluxetable}{cccccc}
\tablecaption{Galaxy properties from the NASA-Sloan Atlas}
\label{tab:prop}
\tablehead { \colhead{NSA ID} & \colhead{$z$} & \colhead{Distance} & \colhead{Resolution} & \colhead{M$_*$} & \colhead{Metallicity} \\
\colhead{} & \colhead{} & \colhead{[Mpc]} & \colhead{[pc]} & \colhead{[M$_{\odot}$]}&  \colhead{[M/H]} }
\startdata
67333 & 0.0020 & 11.2 & 2.2 & $8.56 \times 10^7$ & -0.41\\
124477 & 0.0073 & 31.6 & 6.1 & $2.29 \times 10^9$ & -0.02\\
88260 & 0.0079 & 34.2 & 6.6 & $1.37 \times 10^8$ & -0.55\\
57867 & 0.0032 & 13.8 & 7.1 & $1.36 \times 10^8$ & -0.60\\
51928 & 0.0098 & 42.5 & 8.8 & $1.27 \times 10^9$ & -0.32\\
124554 & 0.0117 & 50.9 & 9.8  & $6.09 \times 10^8$ & -0.46\\
104881 & 0.0160 & 69.8 & 13.3 & $5.41 \times 10^8$ & -0.07\\
156688 & 0.0183 & 80.0 & 15.3 & $2.26 \times 10^7$ & -0.45\\
\enddata
\tablecomments{NSA IDs, redshifts, and stellar masses are taken from the NASA-Sloan Atlas Version 0. The metallicities are calculated from SDSS spectra using pPXF \citep{cappellari_full_2023}. {The angular resolution given is for F606W images.}}
\end{deluxetable}

\section{Analysis}

\subsection{GALFIT}

To study the galaxy structure of these eight dwarf galaxies, we use the 2D parametric modeling software, GALFIT \citep{peng_detailed_2010}. We aim to decompose each galaxy into {some combination of four} possible components: a disk, a bulge, {small nuclear component, and a} point source. The bulge and disk component are modeled by the S\'ersic  equation. The S\'ersic  equation describes the surface brightness as function of radius and takes the form
\begin{equation}
    \Sigma(r) = \Sigma_e \exp (-\kappa ((\frac{r}{r_e})^{1/n}-1)).
\end{equation}
Where $n$ is the S\'ersic  index and controls the distribution of the surface brightness. $r_e$ is the effective radius which contains 50 \% of the light. $\kappa$ is defined by
\begin{equation}
    \gamma(2n; \kappa) = \frac{1}{2} \Gamma(2n)
\end{equation}
where $\gamma$ and $\Gamma$ are the incomplete Gamma function and the Gamma function respectively. $\Sigma_e$ is the surface brightness at the effective radius.
A classical bulge is defined by the light profiles of elliptical galaxies and corresponds to a S\'ersic  index $n > 2$, while a disk corresponds to S\'ersic  index $n=1$. Unresolved point sources are modeled with an point spread function (PSF).  A PSF is an empirical 2D model that estimates the spreading of light from a point source on a telescope. We used the Tiny Tim software routine to create PSFs which accounts for the breathing and spatial distortions of \textit{HST} {\citep{krist_20_2011}}. As AGN are unresolved sources at this resolution scale, the PSF estimates the light contribution from the AGN.

We model the different components of each galaxy using the F110W imaging, which is the most sensitive of the three filters to the underlying stellar population and is less impacted by dust. {We created a set of models for each galaxy which included various combinations of S\'ersic components, a psf, and a small nuclear component. For two irregular galaxies (88260, 104881) that are not well described by a simple elliptical shape described by the S\'ersic equation, we enabled fourier and the disk/box modes which create perturbations to the simple S\'ersic equation. These modes can better capture the irregular shapes of the dwarf galaxies that are disturbed by on going star formation.} 
To determine the best fit for each galaxy, each galaxy was fit with four different models composed of different structures. We fit a single component model, a two component model, a single component model with a point source, and a two component model with a point source. S\'ersic indices were left unconstrained in single component models. We also tested to see if a small nuclear S\'ersic component provided a better fit with and without a point source included. 

 {We choose the best fit model with the Akaike Information Criterion (AIC). AIC penalizes model choice based on the number of free parameters in the model. AIC is defined as}
 \begin{equation}
    AIC = 2k - \log(\chi^2_\nu)
\end{equation}
{where $k$ is the number of components included in the GALFIT model.
 GALFIT produces an internally generated $\chi^2_\nu$. However, this number is calculated over entire 2D image. Background sources such as nearby stars and galaxies mean that small changes to the galaxy profile, such as the inclusion of a PSF, do not significantly impact $\chi^2_{\nu}$ generated by GALFIT. We calculate $\chi_\nu$ of a 200 $\times$ 200 pixel cutout of the galaxy. This eliminates background sources from the image that were not modeled. We then use the parameters and number of components from F110W model to initialize the model for the F606W filter.}

\begin{figure*}
    \centering
    \includegraphics[trim={8cm 0 8cm 0},clip,width=0.7\textwidth]{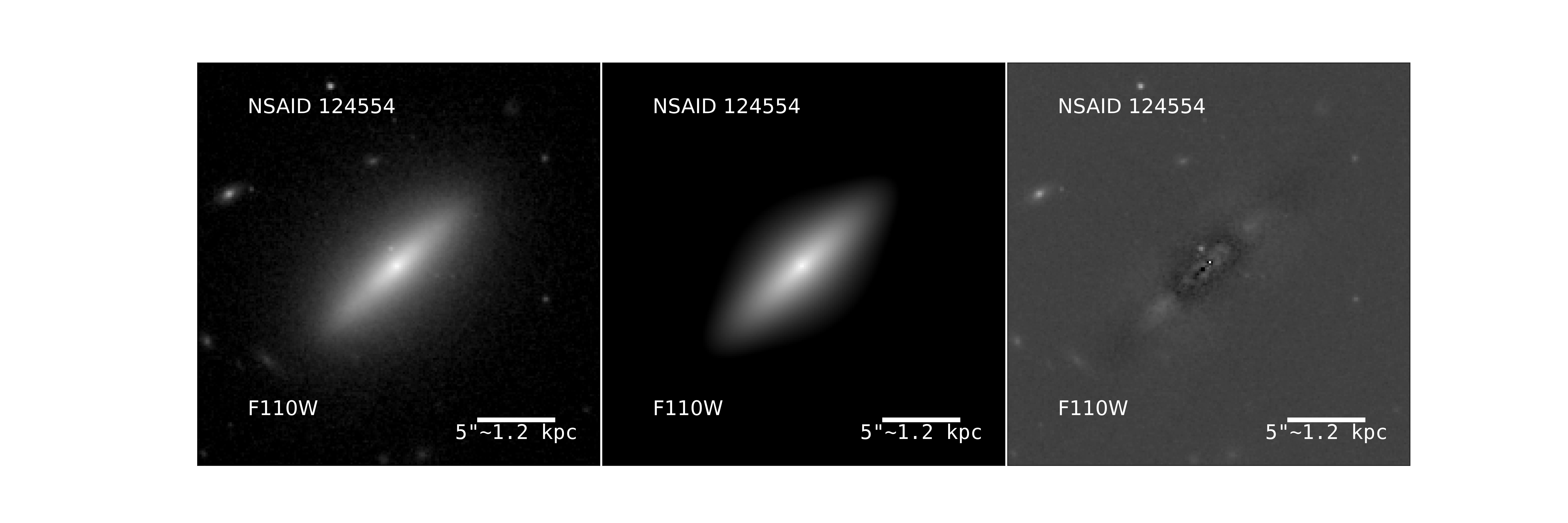}
    \includegraphics[width=0.25\textwidth]{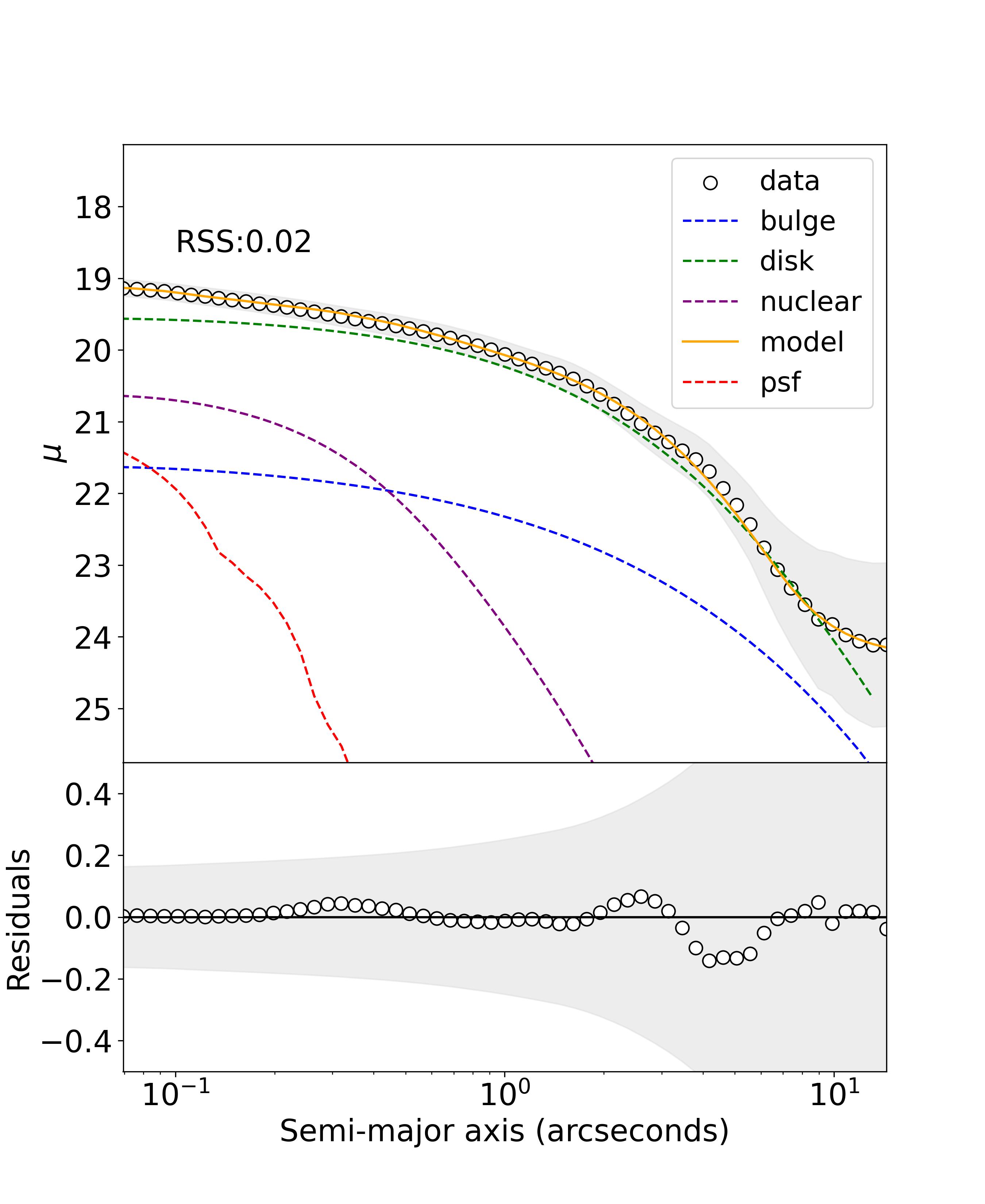}\\

    \caption{Left Panel: Image from HST of NSA 124554. Middle: Model generated with GALFIT. Right: Residual image after model is subtracted from the data. The radial profile is shown on the far right. The orange line shows the model surface brightness profile and the black circles show the surface brightness profile extracted from isophotal fitting. The dashed lines represent the different components of the model. The blue line corresponds to the (pseudo)bulge component. The green line is the disk component. The red line is the PSF and the purple dashed line is a nuclear component. In the bottom panel, we show the residuals between the data and the model.}
    \label{fig:124477-2D}
\end{figure*}

The results of the fitting are listed in Table \ref{tab:galfit-results}. For the irregular galaxies, we fit a single S\'ersic component to get a rough estimate of the underlying galaxy contribution. Figure \ref{fig:124477-2D} shows an example of the best-fit model created by GALFIT. We also utilized Photutils Isophotal fitting \citep{larry_bradley_astropyphotutils_2024} to create 1D profiles of the models as shown in Figure \ref{fig:124477-2D}, for comparison to the GALFIT results. 
We were able to obtain fits for all galaxies except NSA 156688, which is edge-on. 
More details on the individual galaxies can be found in the Appendix {and Figures \ref{fig:galaxy-fits-1} and \ref{fig:galaxy-fits-2}}. 

The error on the S\'ersic components' magnitude is largely due to uncertainties in the background. The uncertainty of the background was calculated by iteratively subtracting pixels above five sigma and taking the standard deviation of the resulting background value. The errors of the magnitude of the point source are those reported by GALFIT. 
The largest source of uncertainty on the values of S\'ersic  index $n$ and effective radius $R_e$ is the choice of PSF model. A poor PSF model can result in an inaccurate model overall. We created a secondary PSF from bright stars in the images themselves to generate uncertainties on the model parameters. Ideally, when using stars to model the PSF, there is a star close to the center of the observation to control for spatial variations across the telescope and there is star for every observation made. These conditions are not met in our observations. Across the eight observations, there are seven bright stars that we can use to create a secondary PSF by creating star cutouts of the stars and averaging them. The difference between the values of $n$ and $R_e$ obtained with the different PSF models is the uncertainty on these values. 

All the galaxies were best fit with pseudobulges with a median value of $n=1.7$ rather than classical bulges, as is commonly observed in dwarf galaxies \citep{kimbrell_diverse_2021, amorin_host_2009, schutte_black_2019}. Three galaxies are fit with both a bulge and a disk. The remaining four galaxies are fit with a single bulge component. Of the seven successfully fitted galaxies, four of them are fit with point sources. The two galaxies that are not fit with point sources are star forming and have no clear photometric center. These results are listed in Table \ref{tab:galfit-results}. These galaxies are known to be variable between 0.01 - 0.04 magnitudes. This primarily should affect the magnitude of the point source, but the variations are similar to the systematic errors reported by GALFTT, so it is unlikely to strongly affect the overall fits determined by GALFIT. 


\begin{deluxetable*}{cccccccccccccc}
\rotate
\label{tab:galfit-results}
\caption{GALFIT fitting results for the F110W filter. }

\tablehead{\colhead{NSA ID} & \colhead{$R_{e, bulge}$} & \colhead{$n_{bulge}$} & \colhead{$m_{st, bulge}$} & \colhead{$m_{ab, bulge}$} & \colhead{$R_{e, disk}$} & \colhead{$n_{disk}$} & \colhead{$m_{st, disk}$} & \colhead{$m_{ab, disk}$} & \colhead{$R_{e, nuclear}$} & \colhead{$n_{nuclear}$} & \colhead{$m_{st, nuclear}$} & \colhead{$m_{ab, nuclear}$} & \colhead{PSF}  \\ 
\colhead{} & \colhead{pc} & \colhead{} & \colhead{} & \colhead{} & \colhead{pc} & \colhead{} & \colhead{} & \colhead{} & \colhead{pc} & \colhead{} & \colhead{} & \colhead{}  & \colhead{}}
\startdata
67333 & 278(5) & 2.52(0.09) & 15.92(0.05) & 14.30(0.05) & - & - & - & - & - & - & - & - & 21.90(0.02)  \\
124477 & 308(20) & 0.93(0.16) & 17.53(0.75) & 15.47(0.75) & 2363(7) & 1.0 & 14.72(0.06) & 13.11(0.06) & 51(31) & 2.82(1.18) & 18.93(2.71) & 17.31(2.71) & -   \\
88260 & 716(6) & 1.63(0.02) & 17.37(0.09) & 15.75(0.09) & 231(5) & 0.92(0.05) & 19.23(0.52) & 17.61(0.52) & - & - & - & - & -  \\
57867 & 564(9) & 1.19(0.01) & 18.54(0.09) & 16.92(0.10) & - & - & - & - & - & - & - & - & -  \\
51928 & 502(2) & 1.19(0.02) & 16.50(0.05) & 14.88(0.06) & 973(19) & 0.92(0.01) & 16.71(0.06) & 15.09(0.06) & - & - & - & - & 21.97(0.06)  \\
124554 & 1379(26) & 1.37(0.01) & 17.76(0.08) & 16.14(0.08) & 718(1) & 1.0 & 17.58(0.07) & 15.97(0.07) & 87(21) & 0.58(0.56) & 21.25(1.97) & 19.64(1.97) & 24.20(0.20)  \\
104881 & 2061(2) & 1.64(0.10) & 17.84(0.12) & 16.22(0.12) & 638(30) & 1.0 & 18.20(0.16) & 16.58(0.16) & - & - & - & - & 23.12(0.05)  \\
156688 & 4059(6) & 0.55 (0.01) & 15.85(0.04) & 14.23(0.04) & 723(35) & 1.0) & 18.43(0.40) & 16.82(0.40) & - & - & - & - & 22.50(0.03) \\ 
\enddata
\tablecomments{Columns (4) \& (8) is magnitudes reported in STmag system. 
Columns (5) \& (9) is magnitudes reported in the AB magnitude system.
Column (10) is reported in the STmag system.}
\end{deluxetable*}




\subsection{Aperture Photometry}
 As a compliment to GALFIT, we perform aperture photometry on the brightest point source in each host galaxy. For objects where a PSF was included in the fit, the brightest point source corresponds to the PSF. This is both a check on the estimates made by the parametric fitting and a way to estimate the brightness of sources that are visually present, but were not successfully fit by GALFIT. Parametric models such as GALFIT are favored for measuring the light from point sources, since it is difficult to disentangle the underlying galaxy emission and the source. Following the method outlined by \cite{greene_black_2008}, we measure {the flux within an aperture with a radius of}  three pixels with a background annulus with an inner radius of 5 pixels and an outer radius of 10 pixels to estimate the local background. We then scale the Tiny Tim PSF model used in the parametric fitting process to match the measured flux in that aperture. This method assumes that all the light from the aperture is from the AGN, hence it will likely overestimate the AGN luminosity.

\begin{figure*}
    \centering
    \includegraphics[width=1.0\textwidth]{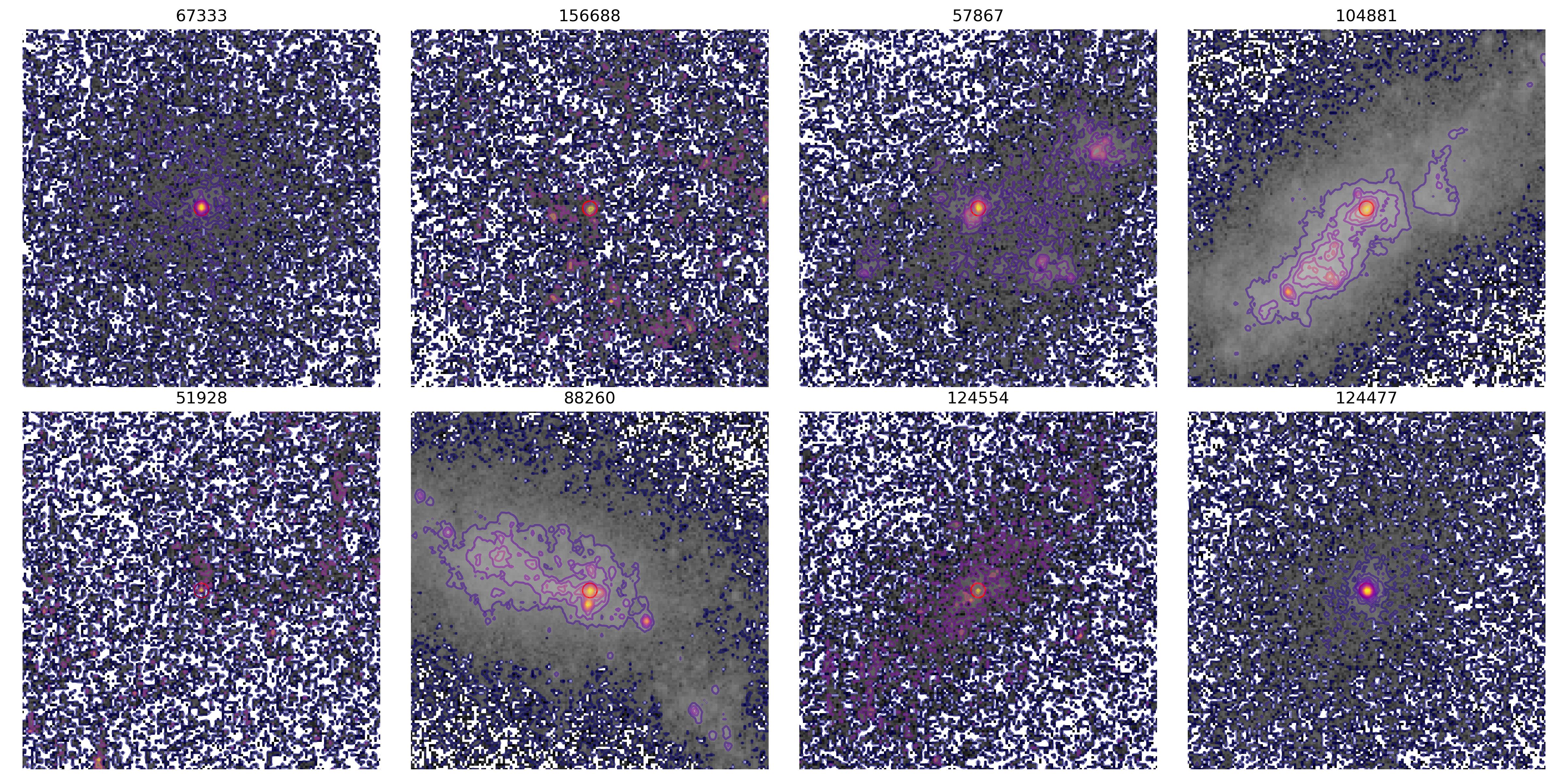}
    \caption{Images from the F300W filter, with the brightest point encircled in red in each image. We overplot the isointensity contours on the image. NSA 51928 does not have any significant UV emission. Images are 14$''$ on each side.}
    \label{fig:300}
\end{figure*}

For the F300X filter, the emission is too diffuse or dominated by the emission from individual star clusters to model with GALFIT. Thus aperture photometry is the only way to estimate point source luminosities in the UV. Figure \ref{fig:300} shows the UV images of the galaxies. 4/8 have a single nuclear source (67333, 124477, 124554, 156688). 3/8 have  both bright point sources and extended UV emission consistent with young stellar populations (57867, 104881, 88260). 1/8 does not have any significant UV emission (51928).

We correct for galactic dust using the Balmer decrement calculated from SDSS spectra to estimate the intrinsic flux value \citep{calzetti_dust_2000}.
The aperture values and their corrections are listed in Table \ref{tab:aperture-phot}. In F110W and F606W, the aperture measurements are systematically brighter than those estimated with GALFIT. Several of the galaxies' UV/optical emission is heavily impacted by dust - in particular 156688 and 51928 - with corrections of up to 4-5 magnitudes in the optical. 

\begin{deluxetable*}{ccccccccc}
\label{tab:aperture-phot}
\caption{Aperture Photometry}

\tablehead{\colhead{} & \colhead{67333} & \colhead{124477} & \colhead{88260} & \colhead{57867} & \colhead{51928} & \colhead{124554} & \colhead{104881} & \colhead{156688}} 
\colnumbers
\startdata
\cutinhead{F110W}
PSF & 21.90(0.04) & - & - & - & 21.97(0.06) & 24.20(0.20) & 23.12(0.05) & 22.50(0.03)\\
PSF Corrected & 21.90(0.04) & - & - & - & 21.03(0.08) & 24.20(0.20) & 22.58(0.05) & 21.47(0.47) \\
Aperture & 21.07(.04) & 20.62(0.09) & 22.94(0.05) & 25.20(0.02) & 20.99(0.02) & 22.06(0.03) & 22.00(0.03) & 25.98(0.04) \\
Corrected & 21.07(0.04) & 20.62(0.09) & 22.47(.06) & 24.78(0.02) & 20.04(0.06) & 22.06(0.03) & 21.46(0.03) & 24.95(0.47) \\
\cutinhead{F606W}
PSF & 20.76(0.02) & - & - & - & 22.31(0.04) & 23.17(0.04) & 22.32(0.05) & - \\
PSF Corrected & 20.76(0.02) & - & - & - & 19.21(0.28) & 23.17(0.05) & 21.19(0.06) & - \\
Aperture & 20.27(0.03) & 19.46(0.02) & 22.11(0.01) & 23.28(0.02) & 23.47(0.03) & 21.97(0.02) & 21.49(0.01) & 25.17(0.03) \\
Corrected & 20.27(0.03) & 19.36(0.02) & 21.34(0.06) & 22.75(0.6) & 20.36(0.28) & 21.97(0.2) & 20.37(0.04) & 21.66(2.35) \\
\cutinhead{F300X}
Aperture & 22.25(0.03) & 21.02(0.02) & 20.78(0.01) & 22.08(0.01) & 28.20(0.70) & 23.85(0.02) & 20.19(0.1) & 23.67(0.02)\\
Corrected & 22.25(0.03) & 21.02(0.02) & 19.53(0.13) & 21.40(0.14) & 21.71(0.84) & 23.84(0.02) & 18.17(0.09) & 16.27(5.29) \\
\enddata
\tablecomments{All magnitudes are reported in the STmag system.
The error reported is systematic error due to variations in the PSF}
\end{deluxetable*}

\subsection{Stellar Population Models \label{sec:ssps}}

To investigate the nature of the emission from the bright source we compare our measurements to stellar population models. While all the galaxies analyzed have some evidence of AGN activity, we want to pay special attention to the bright central sources in these galaxies to determine the origin of the point source. We use stellar population models to characterize the colors, luminosities and masses of the sources. 


Following \cite{kimbrell_diverse_2021}, we used Starburst99 \citep{leitherer_starburst99_1999} to produce model spectra with solar metallicity and a Kroupa IMF for a stellar cluster based on the nuclear star cluster mass predicted by the total stellar mass \citep{neumayer_nuclear_2020}. The model spectra was then convolved with the F110W HST band to produce realistic simulated observations of a star cluster using \texttt{synphot}. We compare the luminosity of the point source to the simulated observations to test whether the luminosity measured is consistent with stellar populations or AGN. 

To further constrain the nature of the point sources, we test whether their colors are consistent with the color evolution of star clusters. To compute colors, we use models from \cite{worthey_comprehensive_1994} which use synthetic spectra that are then are convolved with the specific \textit{HST} filters used in this work. We used a Kroupa IMF, $[M/H]=0.0$ and solar and subsolar metallicity, $[M/H]=-0.33$. It is important to note that synthetic UV spectra performs worse when compared to observations and can produce varaiations in the color of 0.2 magnitudes. 

As another test, we use stellar population models to estimate the stellar mass of the point sources assuming all the light is from stars. We assume that all the light from the central source is from stars and following the method outlined in \cite{georgiev_masses_2016}, which uses stellar population models from \cite{bruzual_stellar_2003}. These models produce color dependent mass to light ratios in WFC3 filters, removing the need to do any conversion between filter systems. We calculate the stellar mass of the brightest source using solar metallicity and a Kroupa IMF using the dust corrected values from both our PSF measurements and aperture photometry. The mass estimates will vary by a factor of two due to assumption of metallicity and stellar age \citep{georgiev_masses_2016}.

\section{Results}

\subsection{Mass Estimates}

Bulge mass is correlated to BH mass much like total stellar mass and we can utilize this scaling relation to estimate BH masses with the fitting results from GALFIT. The scaling relation between bulge mass and BH mass has been extended down into the low mass regime to $ M_{BH} \approx 10^5 M_\odot$ \citep{schutte_black_2019}. To use this scaling relation, we convert the bulge luminosities to masses using color dependent mass ratios from \cite{zibetti_resolved_2009}.
The conversion is given in the following equation
\begin{equation}
     \log(M/L_J) = 1.398(r-z) -1.271
     \label{eq:colormassratio}
\end{equation}
The color to mass ratio uses models that take into account dust and star formation. 

Following \cite{schutte_black_2019}, we first transform our bulge luminosity measurements in the native \textit{HST} to to 2MASS and SDSS filters. HST F110W filter approximately corresponds to the Johnson J filter although it is not exact.  To convert the HST fluxes to the 2MASS \citep{cohen_spectral_2003} and SDSS wavelengths, we fit a power law in $log(f_\lambda)$ vs $log(\lambda)$ space and evaluate the fitted line at the given filter's pivot wavelength. Stellar population models show that a power law is appropriate fit for stellar populations for wavelengths greater than 4000 \rm{\AA} \citep{leitherer_starburst99_1999}. 
The mass estimates for the components of the galaxy are listed in Table \ref{tab:mass}. We use the transformed magnitudes in Equation \ref{eq:colormassratio} to estimate the stellar mass of the bulge component. {The total stellar masses derived differ from those given by NSA by an average of $\approx 0.28$ dex. Different stellar mass estimates can vary by $0.2$ dex and the scatter in stellar mass estimates increases for low mass galaxies by up to $\approx 0.4$ \citep{reyes_stellar_2024}.} 

\begin{deluxetable*}{ccccc}
    \label{tab:mass}
    \caption{Mass Estimates}
    \tablehead{\colhead{NSA ID} & \colhead{$\log_{10}(\frac{M_{\rm bulge,*}}{M_{\odot}} )$} & \colhead{$\log_{10}(\frac{M_{\rm disk,*}}{M_{\odot}} )$} & \colhead{$\log_{10}(\frac{M_{\rm BH}}{M_{\odot}} )_{bulge}$} & \colhead{$\log_{10}(\frac{M_{\rm BH}}{M_{\odot}} )_{total}$}}
    \colnumbers
    \startdata
    67333 & 7.65 & - & 4.64 & 4.23\\
    124477 & 7.77 & 9.06 & 4.80 & 5.73 \\
    88260 & 7.06 & - & - & 4.44\\
    57867 & 7.30 & - & - & 3.64 \\
    51928 & 8.47 & - & - & 5.46\\
    124554 & 8.03 & 8.39 & 5.11 & 5.12 \\
    104881 & 8.11 & 7.82 & - & 5.07\\
    156688 & - & -  & - & 3.62 \\
    \enddata
\tablecomments{Statistical uncertainties from the stellar mass estimates range from 0.1 - 0.15 dex. Errors on BH mass are 0.68 dex for bulge {scaling relation} and 0.55 dex {due measurement error and intrinsic scatter} for the local scaling relation.}
\end{deluxetable*}

With bulge mass estimated, we calculate BH mass based on the bulge mass scaling relation calibrated to include dwarf galaxies by \cite{schutte_black_2019}:  
\begin{equation}
    \log{M_{BH}/M_\odot} = 8.80 + 1.24\log(M_{bulge,*}/10^{11}M_\odot)
    \label{eq:bulge-scaling}
\end{equation}
We only use this equation on the four galaxies that have regular morphology. For the irregular galaxies, we use the BH mass estimates from the scaling relationship between BH mass and total stellar mass as calibrated  for local AGNs by \cite{reines_relations_2015}. 
\begin{equation}
    \log(M_{BH}/M_\odot) = 7.45 + 1.05\log(M_*/10^{11}M_\odot)
    \label{eq:stellarmass-scaling}
\end{equation}
The BH mass estimates are listed in Table \ref{tab:mass}. We find BH masses in the range $10^{3.6 - 6.6} M_\odot$. We note that there are large systematic uncertainties; Equation \ref{eq:bulge-scaling} has a scatter of 0.68 dex and Equation \ref{eq:stellarmass-scaling} of 0.55 dex {due to measurement error and intrinsic scatter in the relation}. 
 
There is some question of whether the BH mass to bulge relationship continues into the low mass range. 
Many dwarf galaxies have pseudobulges rather than classical bulges \citep{schutte_black_2019, greene_black_2008, jiang_black_2011}, which are theorized to have different physical origins and thus may relate to BH mass in different ways.
Additionally, \cite{jiang_black_2011} finds evidence that the bulge-to-BH mass relation for pseudobulges has a steeper slope than more massive galaxies with classical bulges, while \cite{schutte_black_2019} finds that dwarfs pseudo-bulges continue to scale with BH mass. Because we are extrapolating to lower mass galaxies than in \cite{schutte_black_2019}, these BH mass estimates should be treated with caution. 

\subsection{UV and X-ray Properties}
This sample of galaxies was previously studied with \textit{Chandra} X-ray observations in \cite{messick_x-ray_2023}. They found significant X-ray detections consistent with AGN activity in NSA 104881, 156688, and 51928. The remaining objects were considered weak or non-detections, since the X-ray emission was not significant above the background noise. Combining X-ray and UV observations, we can measure $\alpha_{ox}$, which is a measure of the relative X-ray emission to UV emission. $\alpha_{ox}$ is thought to yield information about the relation between the accretion disk and X-ray corona. It has been shown that AGN in dwarf galaxies are relatively X-ray weak given their UV emission \citep{baldassare_hubble_2017, wasleske_x-ray_2023} with $\alpha_{ox}$ values falling below the relation found \cite{just_x-ray_2007} between $\alpha_{ox}$ and $L_{2500}$ for bright quasars.

$\alpha_{ox}$ is the ratio between $f_{2500}$ and $f_{2kev}$ following the equation
\begin{equation}
    \alpha_{ox} = 0.3838 (\log{f_{2500}/f_{2kev}}).
\end{equation}
We convert our photometric measurements from the F300X filter to L2500 by using a power law of $F_\nu \propto \nu^{-0.5}$ \citep{vanden_berk_composite_2001}. For the X-ray, we get a specific flux density by using a power law index of $\Gamma = 2$.

\begin{figure}
    \centering
    \includegraphics[width=0.5\textwidth]{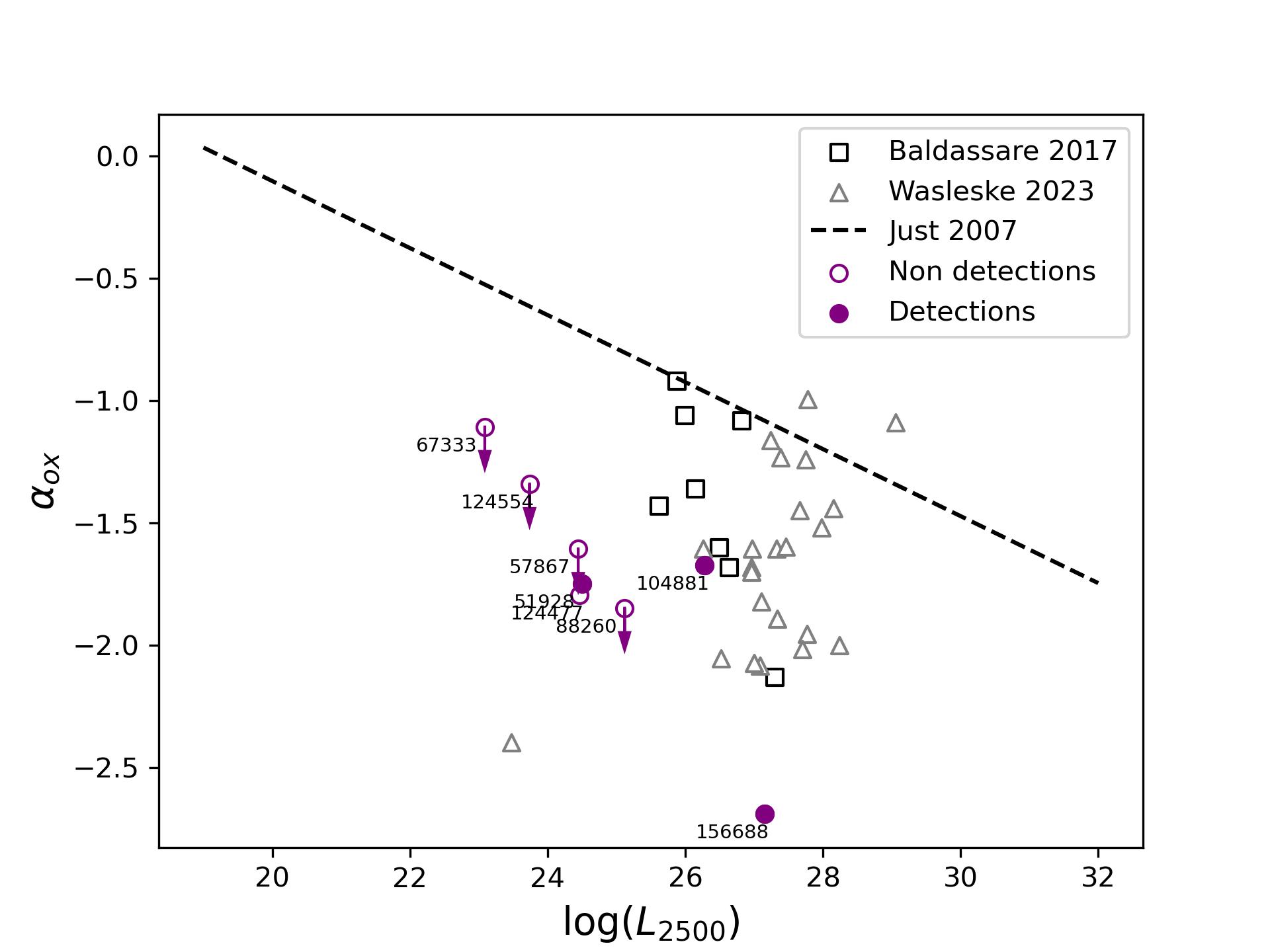}
    \caption{$\alpha$ox plotted against UV Luminosity at 2500 \AA. The purple dots show the points from this work. The black squares show points from \cite{baldassare_hubble_2017}. The grey triangles show points from \cite{wasleske_x-ray_2023}. }
    \label{fig:xray}
\end{figure}

We compare our measurements to \cite{baldassare_hubble_2017} and \cite{wasleske_x-ray_2023} in Figure \ref{fig:xray}. Similar to these samples, our galaxies fall below the relationship from \cite{just_x-ray_2007}, indicating that they are relatively X-ray weak in comparison to bright quasars. Higher accretion rates and lower BH mass are expected to increase $\alpha_{ox}$ \citep{done_intrinsic_2012, dong_x-ray_2012}. This trend is not observed in dwarf galaxies \citep{baldassare_x-ray_2017, wasleske_x-ray_2023}. It is possible that the presence of a star cluster could enhance the UV emission making these objects appear X-ray weak, or that there could be a change in accretion properties at low BH mass.

\subsection{Nuclear Star Cluster or AGN?}
A bright central point source could be an AGN, a nuclear star cluster (NSC), or combination of the two. Nuclear star clusters (NSCs) are dense, compact star clusters with typical effective radii of $< 10 \rm{pc}$ \citep{georgiev_masses_2016}, appearing unresolved at these distances. 
NSCs and MBHs have been found to coexist \citep{seth_coincidence_2008,nguyen_nearby_2018,baldassare_massive_2022}. 
In AGN dwarfs, the point sources are often more luminous than what we would expect to see from an NSC alone \citep{kimbrell_diverse_2021}, distinguishing them from NSCs. 

\begin{figure}
    \centering
    \includegraphics[width=0.5\textwidth]{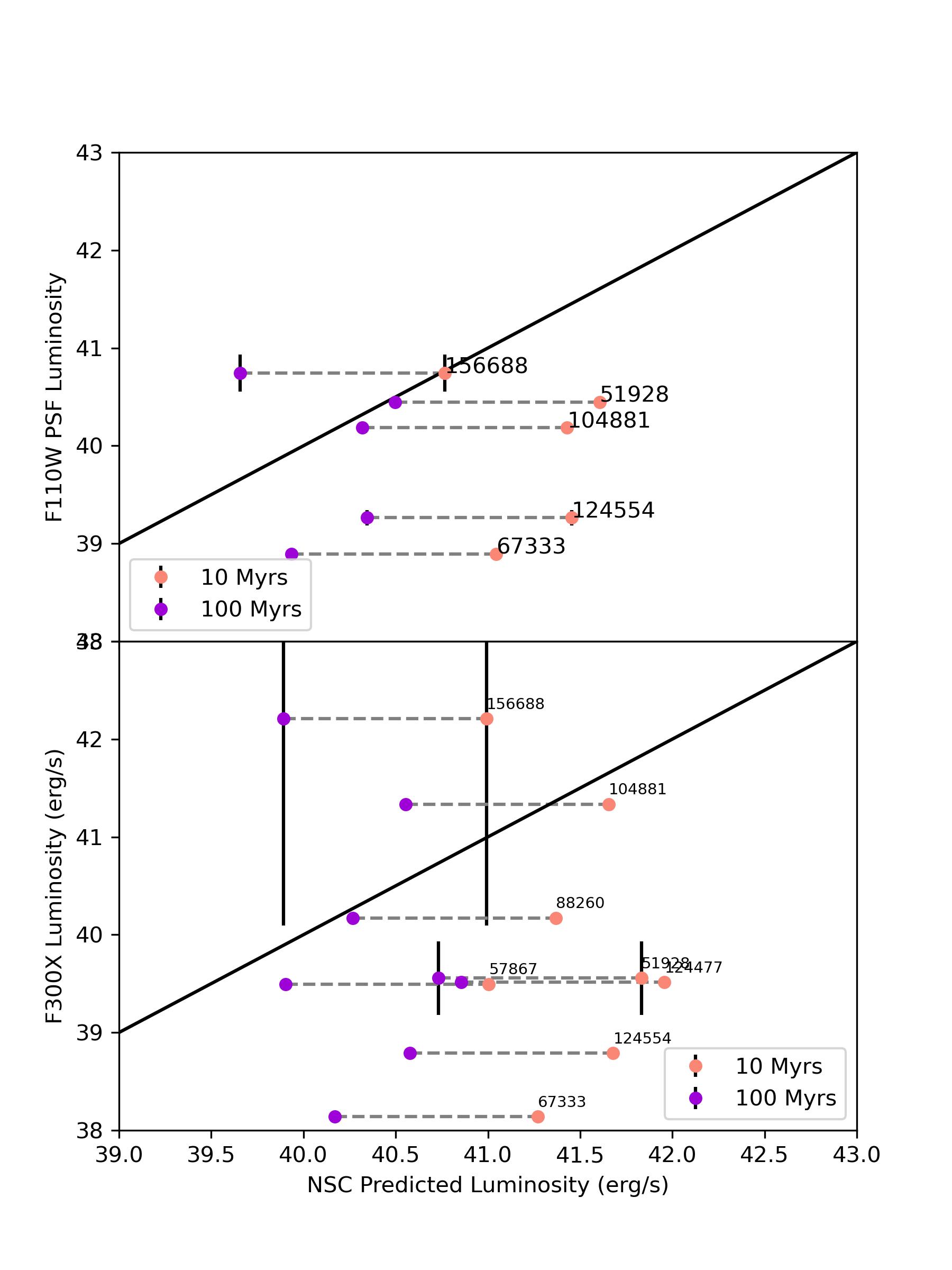}
    \caption{The top figure shows the near-IR PSF luminosity plotted against the predicted luminosity of an NSC. The purple dots show a cluster of 100 Myrs and the orange dots show a cluster of 10 Myrs. The bottom figure, similarly, shows the measured UV luminosity against the predicted NSC UV luminosity. The measured luminosity of the point sources are not significantly brighter than what we would expect for a NSC.}
    \label{fig:nsc-lum}
\end{figure}

To better characterize the nature of the point sources in our galaxies, we compare luminosities and colors of the point source to SSP models from Starburst99 \citep{leitherer_starburst99_1999} and Worthey \citep{worthey_comprehensive_1994} models (see section \ref{sec:ssps}). 
Following \cite{kimbrell_diverse_2021}, we compare the F110W luminosity of the point sources to the emission we would expect to see from a NSC. We utilize the relationship between galaxy mass and NSC mass \citep{neumayer_nuclear_2020} to estimate the mass of NSC.
\begin{equation}
    \log (M_{NSC}) = 0.48 \times \log (\frac{M_*}{10^9 M_{\odot}}) + 6.51
\end{equation}
We then use SSPs from SB99 \citep{leitherer_starburst99_1999} (see section \ref{sec:ssps}). In Figure \ref{fig:nsc-lum}, we show the measured luminosities compared to the synthetic observations for NSCs. Unlike the \cite{kimbrell_diverse_2021} sample, the measured luminosities of our sample are not significantly brighter than what we would expect to see from a NSC. 
    
\begin{figure}
    \centering
    \includegraphics[width=0.5\textwidth]{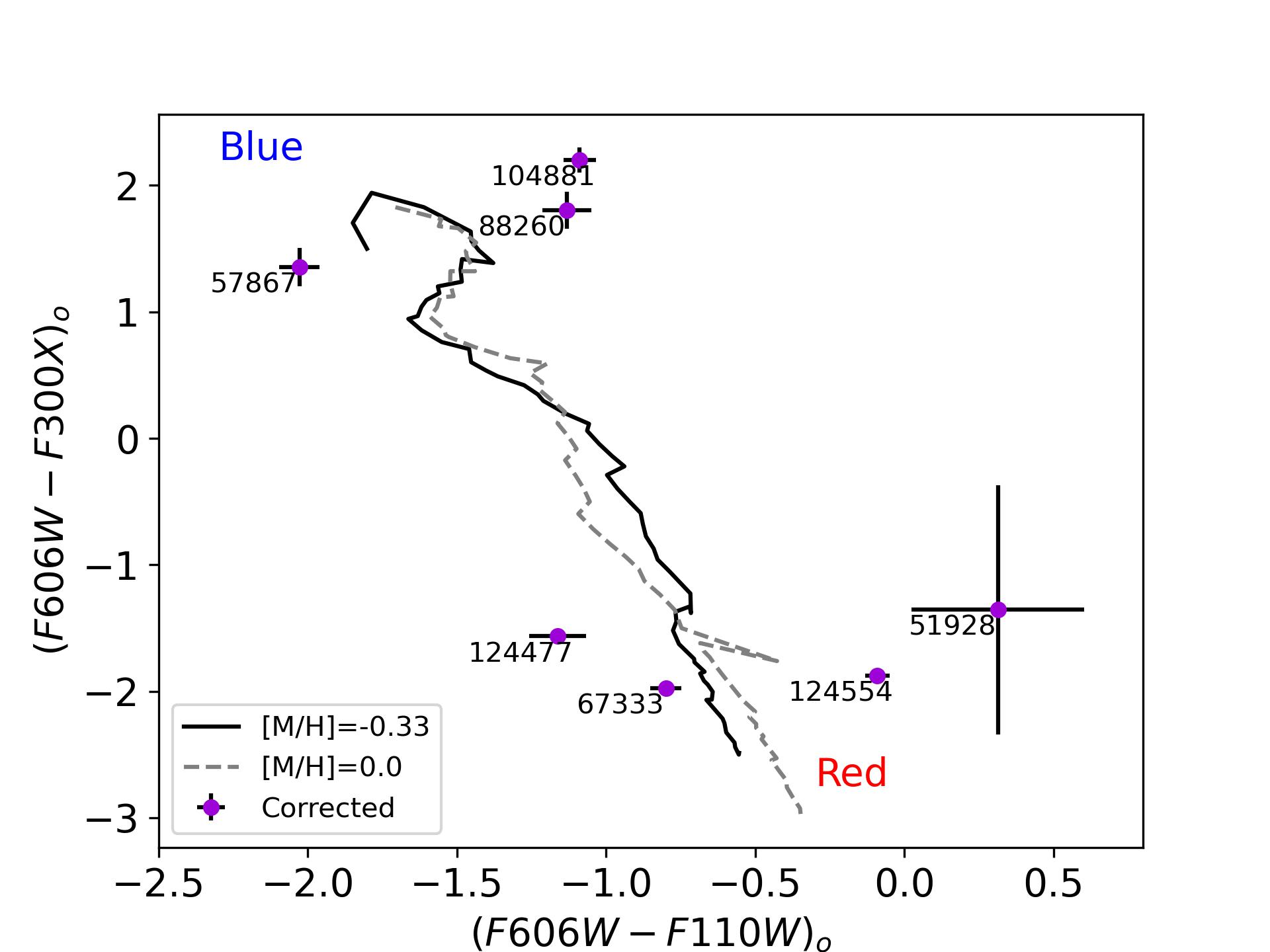}
    \caption{Color-color plot of the brightest point sources. The purple dots show the dust-corrected colors of the point source. The black line and the grey dashed line show SSPs with different metallicities. Five of the galaxies have colors inconsistent with stellar populations. }
    \label{fig:color-color}
\end{figure}

To further put constraints on the dominant source of emission, we use color evolution in conjunction with Worthey SSPs \citep{worthey_comprehensive_1994}.  In Figure \ref{fig:color-color}, we plot the colors of these galaxies along with SSP models \citep{worthey_comprehensive_1994}. {The SSP models track the ages of stellar populations from 11.4 Myrs to 17.8 Gyrs.} 124477, 67333, and 57867 have colors consistent with being star clusters. 57867 has colors consistent with a star cluster that is very young. Five galaxies (88260, 104881, 51928, 156688, 124554) are red in F606W-F110W filter, inconsistent with being star clusters. 
Color selection has also been employed to distinguish between stars, quasars and extended sources \citep{richards_spectral_2006}. The quasars also tend to be redder than other sources, indicating that the colors of the central sources are may be more consistent with AGN activity. 

\begin{figure}
    \centering
    \includegraphics[width=0.5\textwidth]{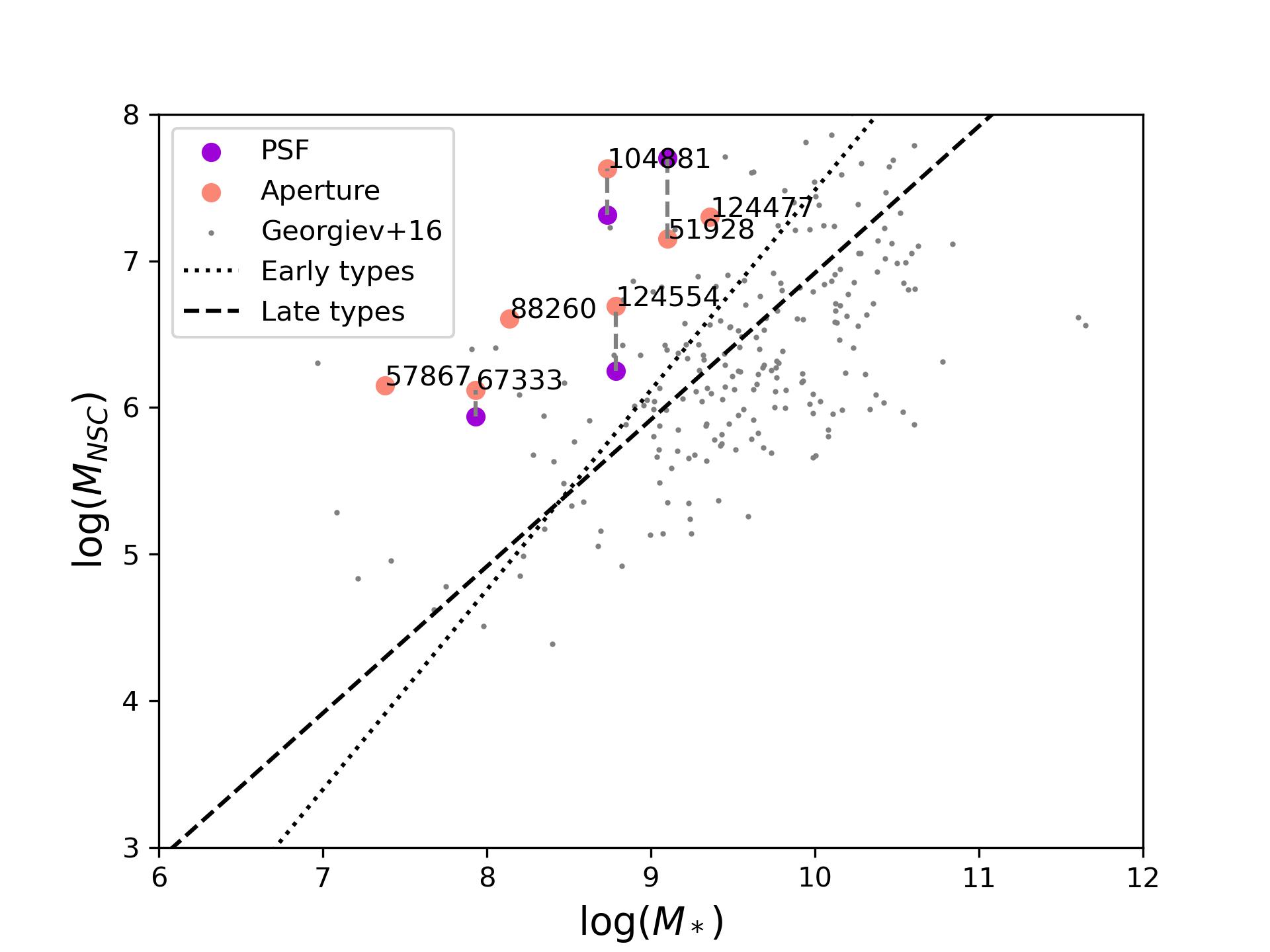}
    \caption{Measured NSC mass given the color of the cluster plotted against total stellar mass. The purple dots show the predicted nsc mass based of its PSF measurements. The orange dots show the predicted NSC mass based on its aperture photometry measurements. The grey dots show the NSC masses measured by \cite{georgiev_masses_2016}. The dashed and dotted lines show the scaling relation between $M_{NSC}$ and total stellar mass for late type galaxies and early type galaxies respectively.}
    \label{fig:nsc-mass}
\end{figure}

Furthermore, we can use the measured color of the point sources to estimate the stellar mass (assuming their emission is primarily stars) using mass to light ratios given by \cite{bruzual_stellar_2003}. Figure \ref{fig:nsc-mass} shows the estimated masses of the clusters plotted against their stellar mass. \cite{georgiev_masses_2016} calibrated scaling relations for nuclear star clusters and their stellar mass. The estimated stellar masses of the point sources are over massive given their total galaxy stellar mass, based on the scaling relations found by \cite{georgiev_masses_2016}. 
These high masses are inconsistent with our previous finding that these point sources are underluminous compared to expectations for NSCs \ref{fig:nsc-lum}. 
While the point sources are not overly luminous for dwarf galaxies of this size, their colors and subsequent stellar mass estimates indicate the emission from these point sources are not entirely dominated by stars.

\section{Discussion}

\subsection{Comparison to Other Samples}

Detailed decomposition of galaxies hosting low mass black holes has been done by others authors \citep{kimbrell_diverse_2021,schutte_black_2019, jiang_host_2011, greene_black_2008, baldassare_hubble_2017}. These samples of AGN have been selected by optical spectroscopy by identifying broad lines and classification through the BPT diagram. The variable AGN dwarf galaxies from this work would not be selected as AGN using these features alone. Optical spectroscopic selection may be biased against low luminosity AGN that are dominated by star formation \citep{trump_biases_2015}. Due to this difference in selection technique, we compare the properties of the variability selected AGN and their host galaxies to these other samples.  

The morphology of our sample is consistent with dwarf AGN hosts found by \cite{kimbrell_diverse_2021}. They found that dwarf hosts are more likely to have pseudobulges and non-AGN host dwarfs were more likely to be disk dominated \citep{kimbrell_comparison_2023}. For the galaxies that are not irregular and star forming we find a similar trend; they host pseudobulges. However, we find that our sample has a higher rate of irregular/star forming galaxies than the \cite{kimbrell_diverse_2021} sample. 

\begin{figure}
    \centering
    \includegraphics[width=0.5\textwidth]{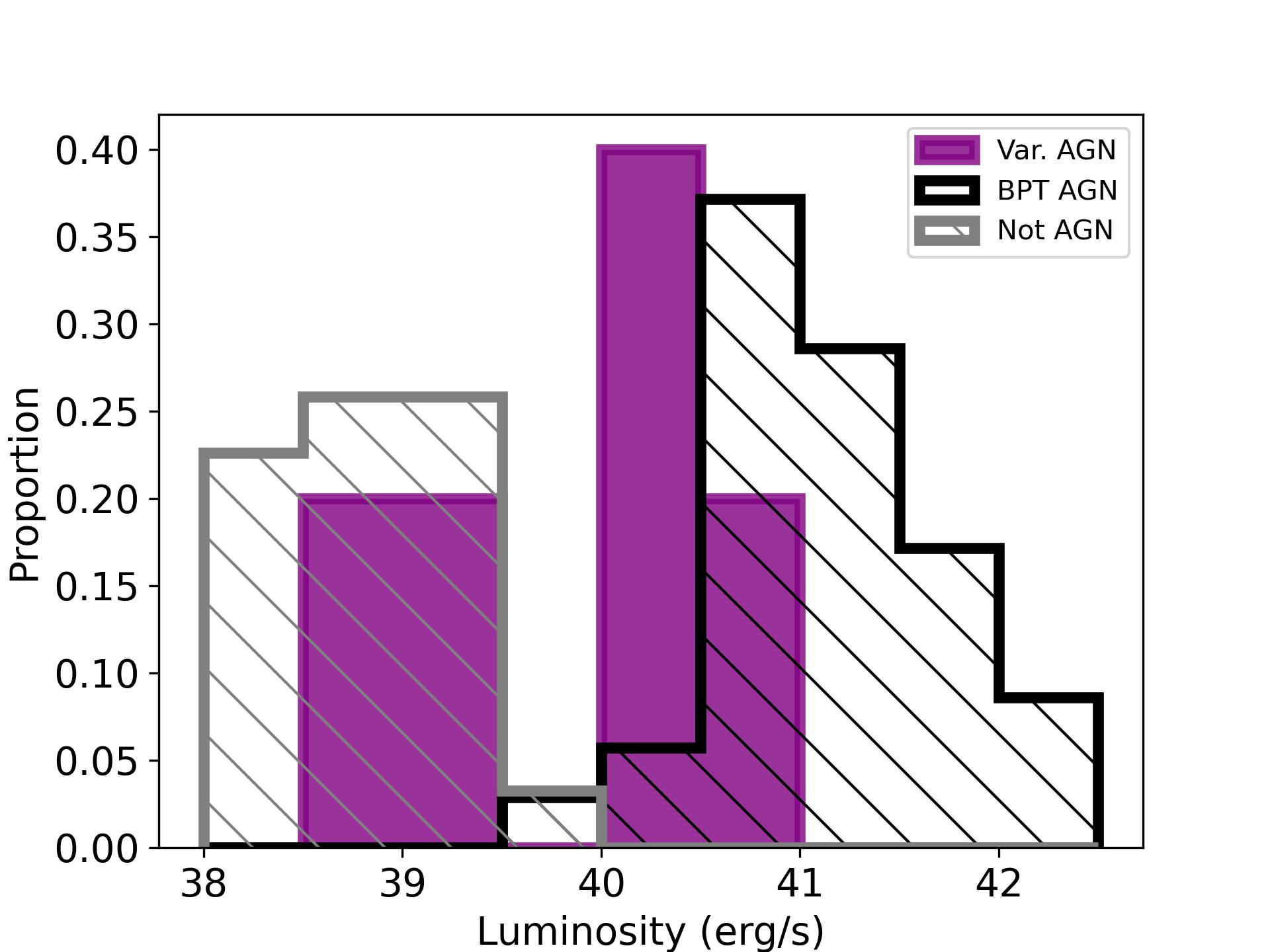}
    \caption{Luminosity of the point source in variability selected AGN (purple),  optically selected AGN (black), and of point sources in dwarfs not identified to be AGN. The point sources in variabilty selected AGN are less luminous than those in optically selected AGN, but more luminous than those in non-AGN host galaxies.}
    \label{fig:agn-lum}
\end{figure}

 Figure \ref{fig:agn-lum} shows the distribution of F110W point source luminosity of variable sample in comparison to BPT selected sample \citep{kimbrell_diverse_2021} and point sources in dwarf galaxies that do not host AGN \citep{kimbrell_comparison_2023}. The point sources of the variable AGN have lower luminosities than the BPT selected AGN, but higher than non-active dwarf galaxies.
On average, the variability selected AGN are approximately 10 times brighter than the non-AGN point sources, but approximately 40 times dimmer than the optically selected dwarfs. This is consistent with what we expect for AGN that are not selected by the BPT diagram. This demonstrates that variability selection may find lower luminosity AGN than methods using optical spectroscopy. 

\begin{figure}
    \centering
    \includegraphics[width=0.5\textwidth]{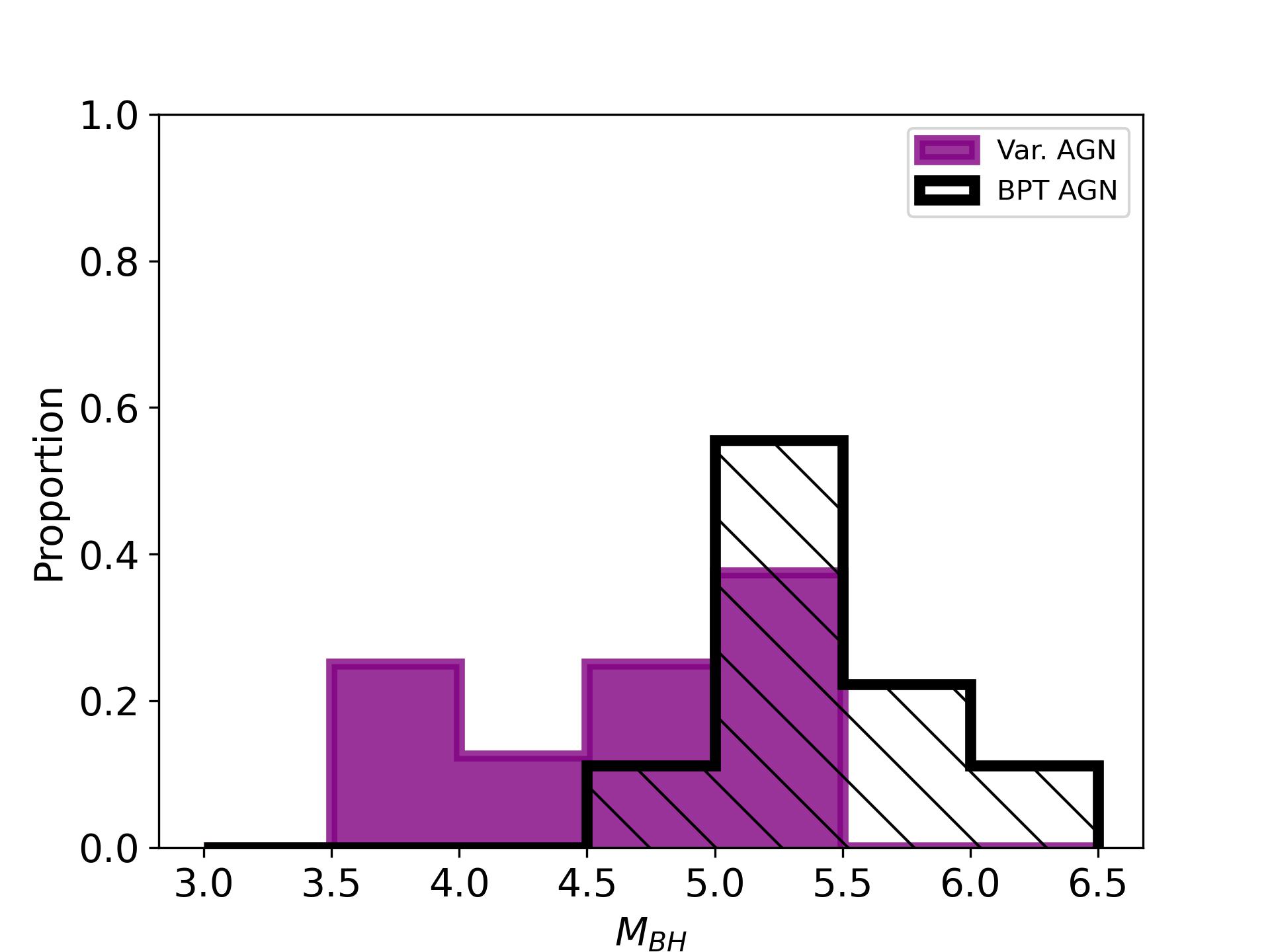}
    \caption{Histogram showing the distribution of estimated black hole masses. The black line shows the average black hole mass from broad-line AGN from \cite{reines_dwarf_2013}. {The error on the BH mass measurements are up to 0.68 dex.} The variable selected AGN in dwarfs have lower masses than broad-line AGN selected by the BPT diagram.}
    \label{fig:bh-mass}
\end{figure}

We also compare the BH masses of the variable sample to \cite{reines_dwarf_2013} whose masses were calculated using broad H$\alpha$ in Figure \ref{fig:bh-mass}. We find that the variable selected AGN have BH masses that are lower than the \cite{reines_dwarf_2013} sample, indicating that variability may be able to find lower mass BHs. It is important to note that while this does indicate that the variable AGN in this sample are in the $10^{3-5} M_\odot$ mass range, the scaling relations we used to estimate the masses have large scatter and the mass estimates are probably only accurate to within an order of magnitude. In addition to the large scatter around these scaling relationships, due to the extreme low stellar mass of some these objects, we have extended the scaling relation past where they have been shown to be true. There is a possibility that the scaling relations change in the low mass regime, depending of the seeding channel. 



\subsection{Optical Spectroscopic Properties}

The BPT diagram is an effective tool for selecting for AGN activity, however it is biased against low mass star forming galaxies \citep{trump_biases_2015} and low metallicity galaxies \citep{cann_limitations_2019}. This sample of galaxies are selected as star forming by the BPT diagram or cannot be plotted on the BPT diagram because they do not have the requisite emission lines. For three of these galaxies (104881, 57867, 88260) it is likely they are impacted by star formation dilution as they have high specific star formation rates (-0.4, -0.3, -0.5 $\log(Gyr^{-1})$, respectively). Their AGN may have little contribution to ionizionation of gas and is dominated by star formation. 
Three of these galaxies do not appear on the BPT diagram at all (67333, 124477, 124554) and have very little evidence of on going star formation. There has been some evidence of dwarf galaxies being gas poor with little evidence of star formation, but still having signatures of ionized gas, falling on the star formation region of the BPT diagram. This could be due to presence of an AGN. These gas depleted dwarfs tend to redder than dwarfs with a normal amount of HI content \citep{bradford_effect_2018}. The dwarfs that do not appear on the BPT diagram in this sample are similar to the gas depleted dwarfs. They have similar colors and have little star formation. It is possible that this subset of galaxies are gas poor which is why they do not appear on the BPT diagram. The remaining two galaxies (156688, 51928) have high Balmer decrements indicating high levels of dust extinction. 

\section{Conclusions}

We obtained \textit{HST} imaging in three different bands - UV, optical and Near-IR - and used GALFIT to characterize the structure of eight dwarf galaxies with variability-selected AGN candidates. We used scaling relations to estimate BH masses and we further analyzed the nature of the point sources in each galaxy to see if they had properties more consistent with a NSC or an AGN. We then compared this sample of variability-selected AGN dwarf galaxies to AGN dwarf galaxies selected via optical spectroscopy. 

\begin{itemize}
  \item The host galaxies of variable selected AGN have diverse structure and properties, with 3/8 of the galaxies having irregular structure. Of the galaxies that were well-modeled with GALFIT, they were all best fit with pseudo-bulges, as is commonly found in dwarf galaxies. 
  \item We estimate BH masses through two scaling relations, the bulge mass to BH mass relationship and the total stellar mass to BH mass relationship. The BH mass estimates range from $10^{3.6-6.6} M_\odot$. We find that the BH mass estimates are, on average, lower than the BH mass estimates for AGNs selected through optical spectroscopy. This shows that variability selection methods can find lower mass BHs, however, more stringent BH mass estimates are needed. 
  \item We investigated the nature of the ``central" point sources in the host galaxies to see if they had properties more consistent with being NSCs or AGN. While these sources are not significantly more luminous than what we would expect from a NSC, 5/8 do not have colors that are consistent with being star clusters. If these point sources are dominated by emission from stars, rather than an AGN, these star clusters are over massive than what we would expect for galaxies of these sizes. We argue that the light from these point sources are likely not entirely dominated by stars. 
  \item We compare the overall AGN luminosity of these objects to two other populations of dwarf galaxies -- AGN dwarfs selected through optical spectroscopy and non-active dwarfs. We find that variability selected AGN are dimmer than the AGN dwarfs selected by optical spectroscopy, but brighter than the non-active dwarfs. This supports the argument that the point sources in this sample are not dominated by stars and shows variability studies can find lower luminosity AGN.
\end{itemize}

We demonstrate that variability-selected dwarf AGNs have lower luminosities and BH masses than AGN selected via optical spectroscopy, but the hosts of variability selected AGN have a diverse range of morphological types, similar to other active dwarfs. 


\begin{acknowledgments}

Support for program HST-GO-16423 was provided by NASA through a
grant from the Space Telescope Science Institute, which is operated by the Associations of Universities for Research in Astronomy, Incorporated, under NASA contract NAS5-26555.

EMK would like to thank {the reviewer for helpful comments and suggestions,} Erik Wasleske for assisting with spectroscopic analysis and Theresa Bunker's 6th graders for insightful comments and discussion.  
\end{acknowledgments}

\bibliographystyle{aasjournal}

\bibliography{biblio}

\appendix 

\section{Notes on Individual Galaxies}

\begin{figure*}
    \centering
    \includegraphics[trim={8cm 0 8cm 0},clip,width=0.7\textwidth]{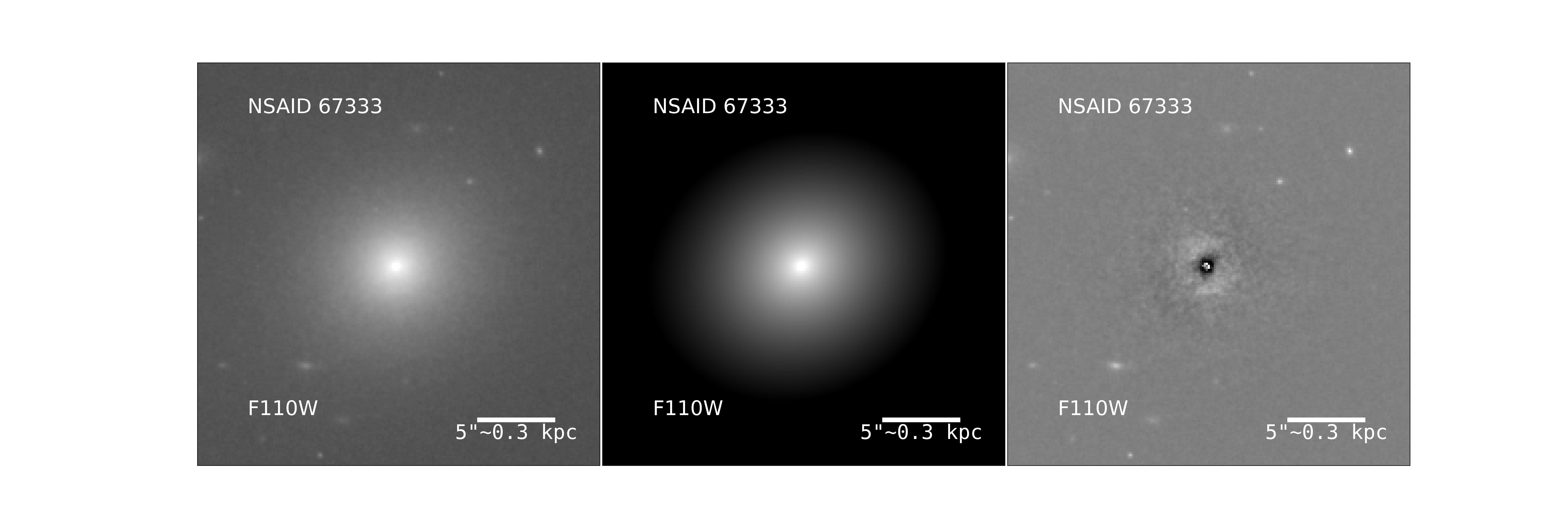}
    \includegraphics[width=0.25\textwidth]{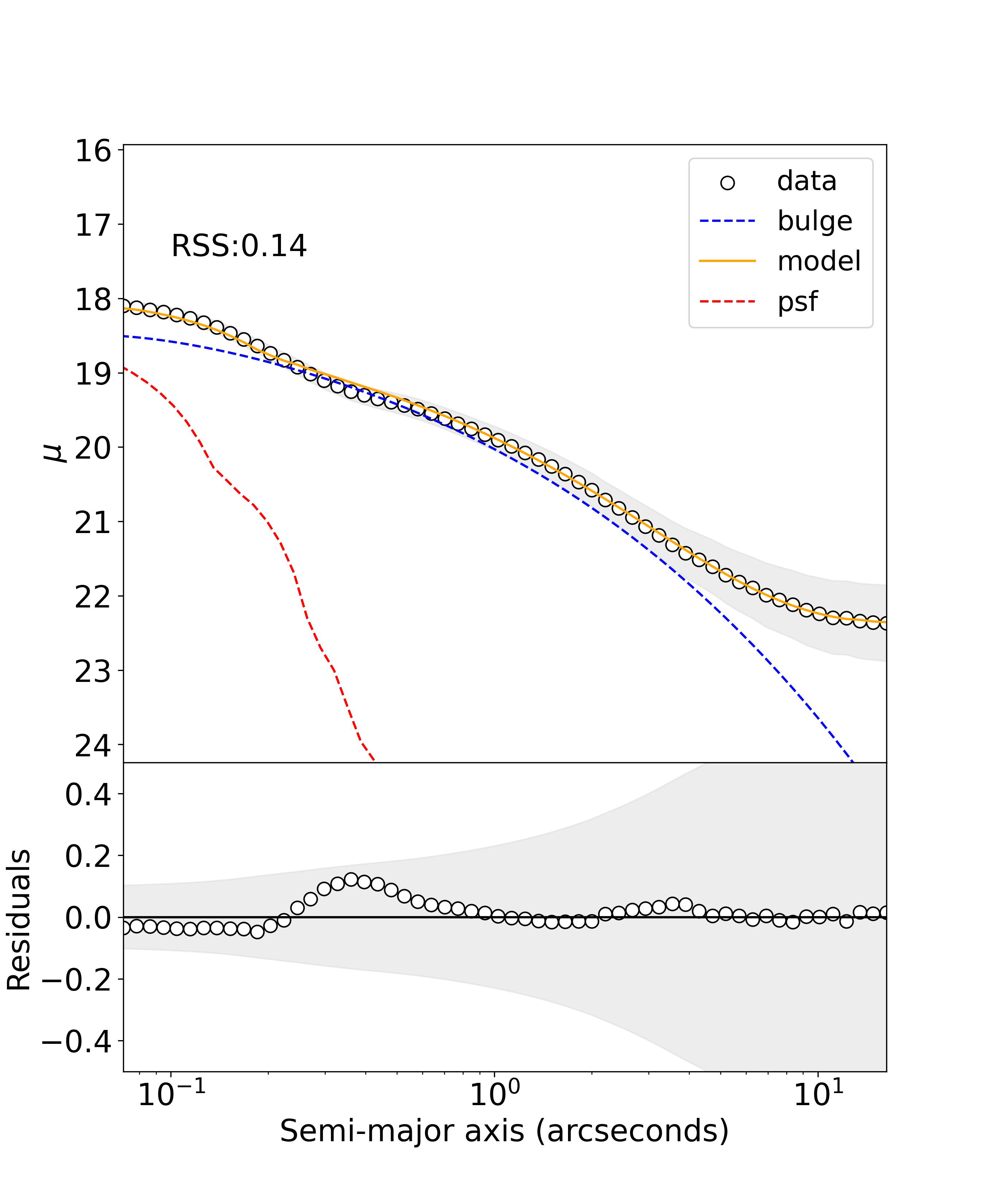}\\
    
    \includegraphics[trim={8cm 0 8cm 0},clip,width=0.7\textwidth]{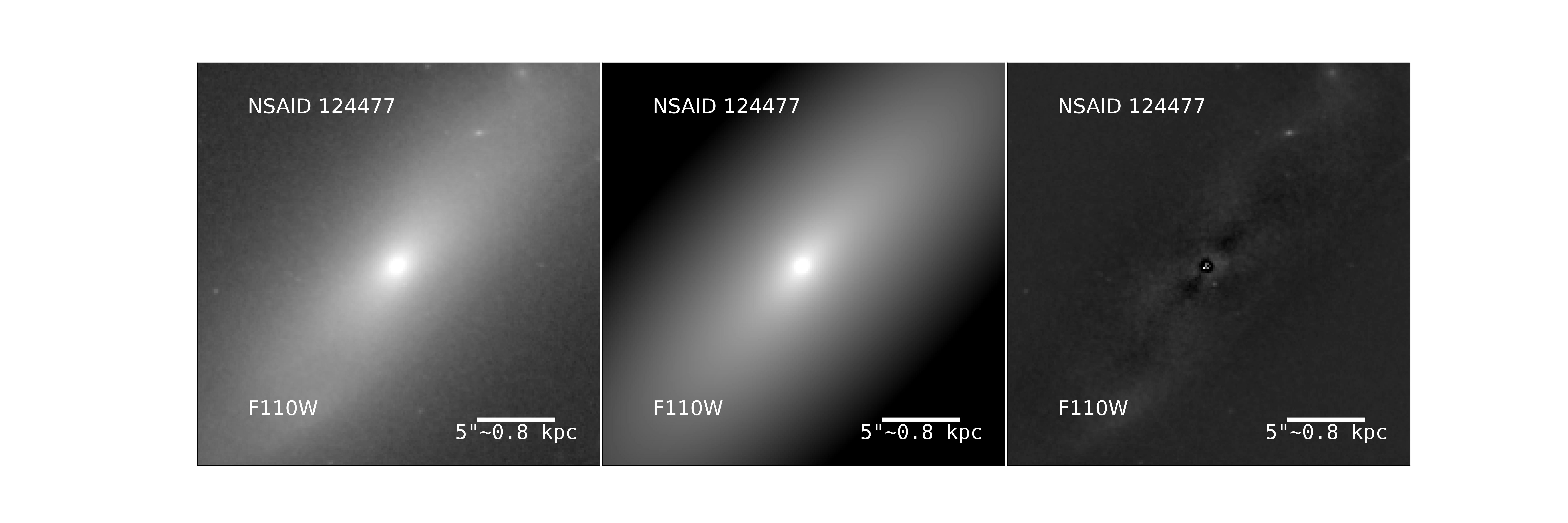}
    \includegraphics[width=0.25\textwidth]{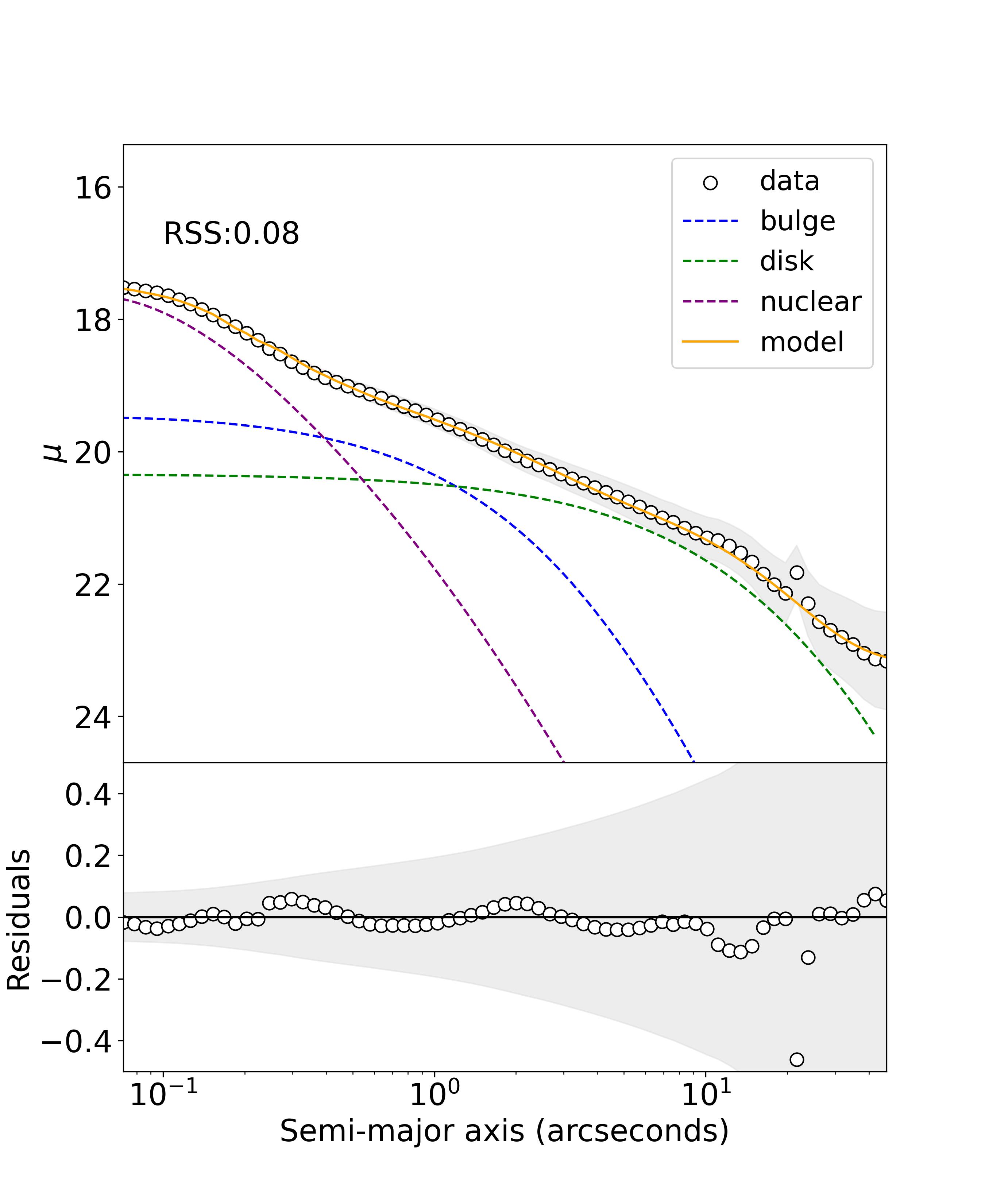}\\

    \includegraphics[trim={8cm 0 8cm 0},clip,width=0.7\textwidth]{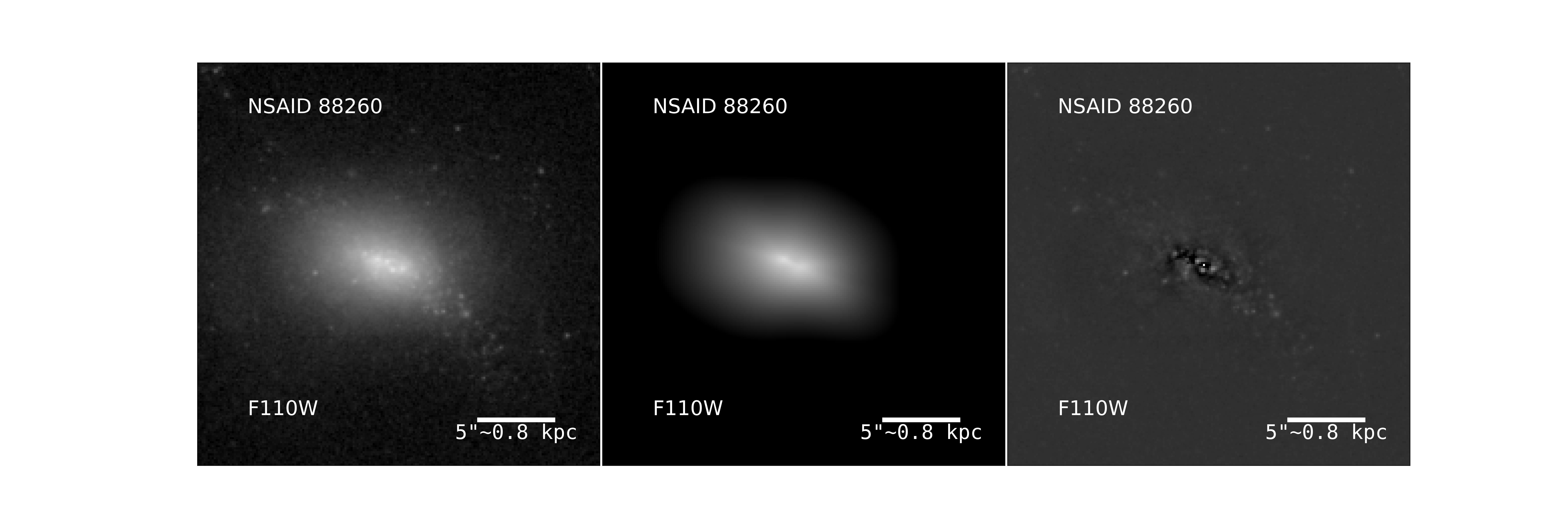}
    \includegraphics[width=0.25\textwidth]{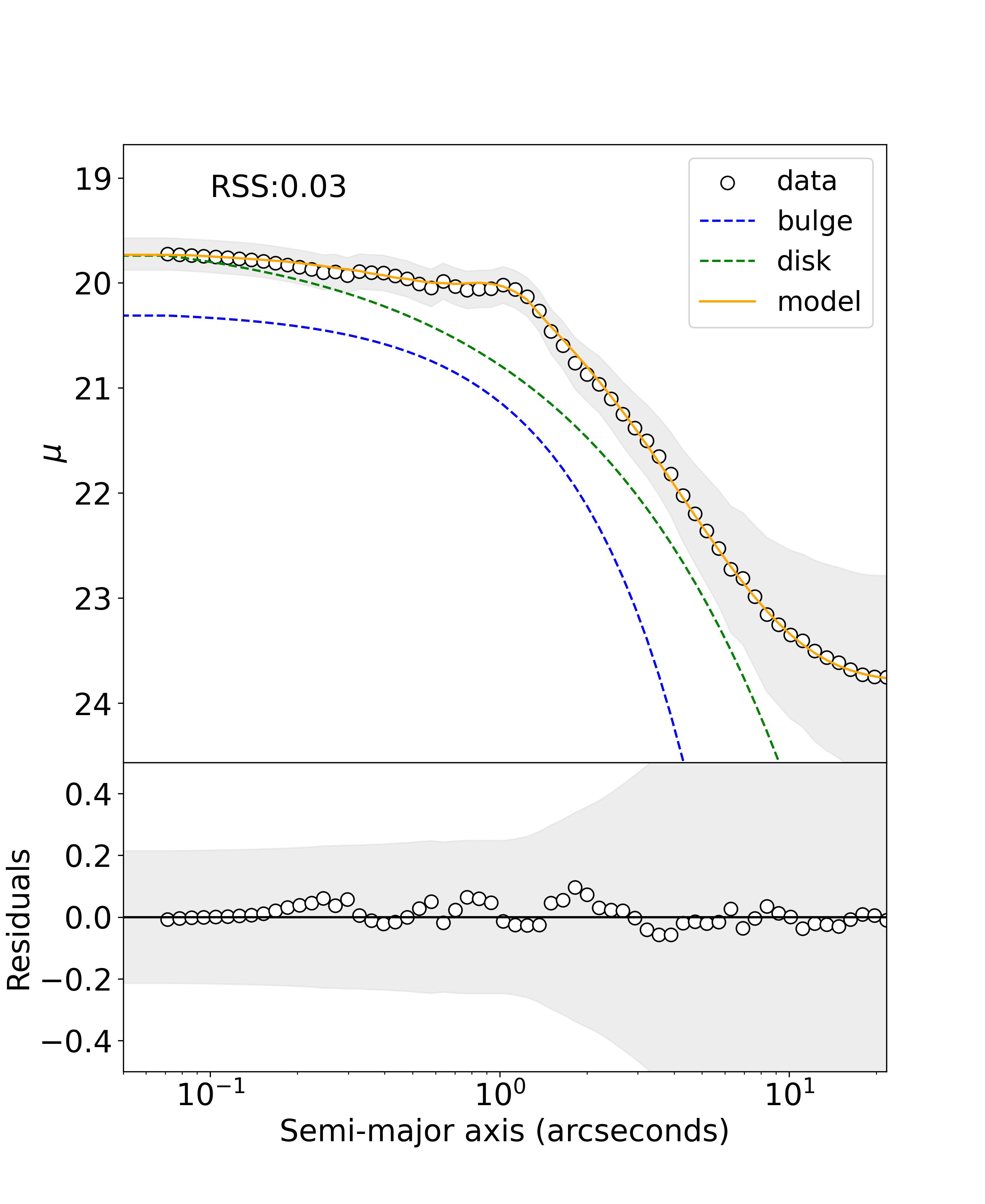}\\

    \includegraphics[trim={8cm 0 8cm 0},clip,width=0.7\textwidth]{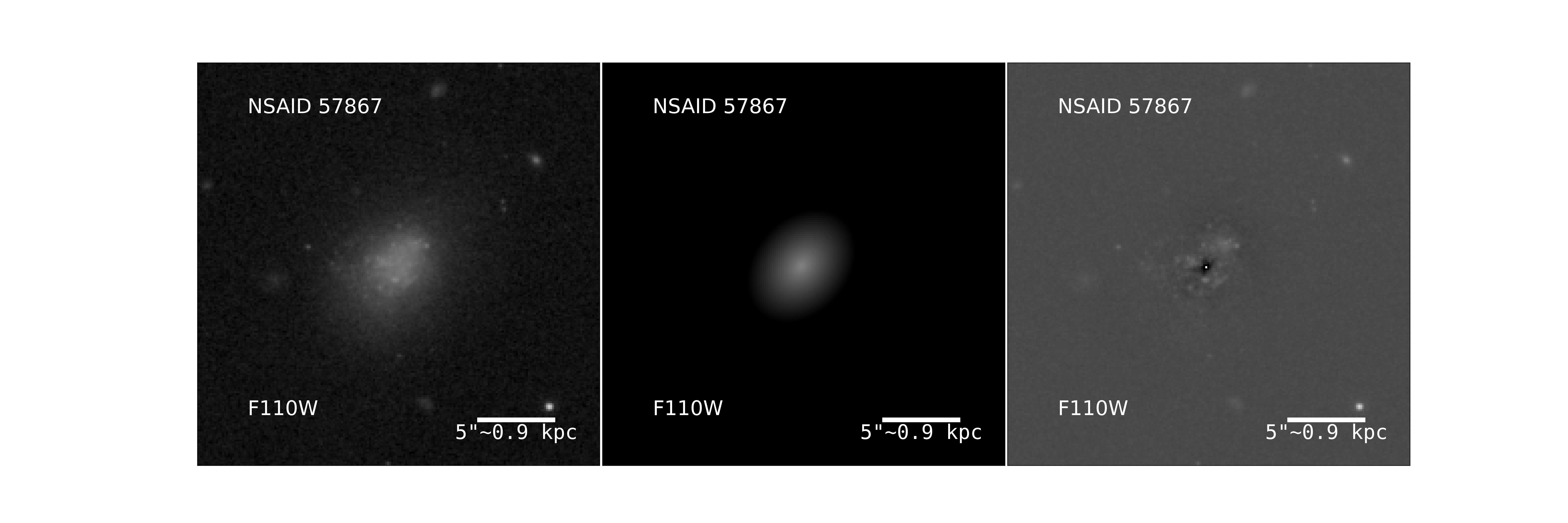}
    \includegraphics[width=0.25\textwidth]{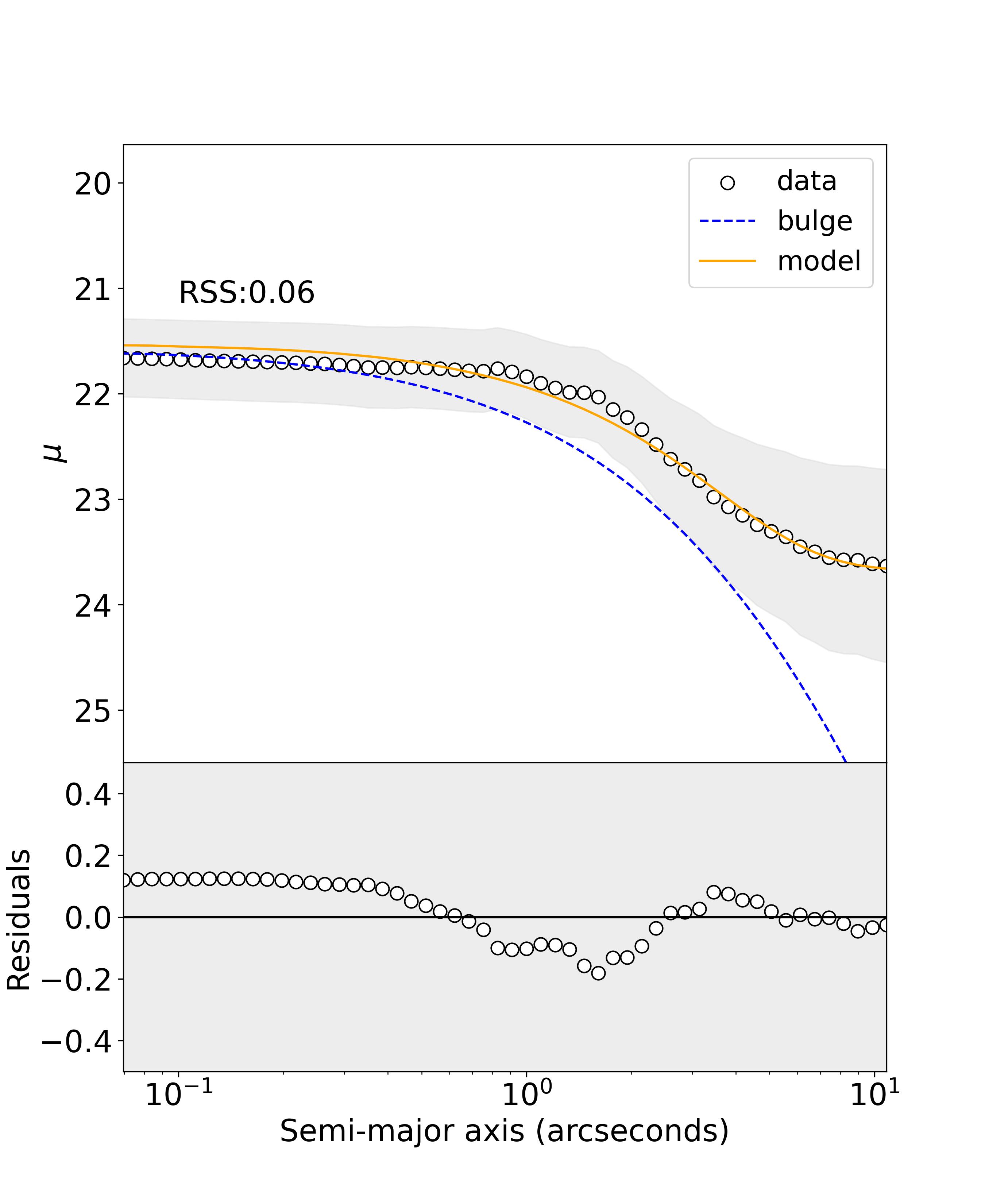}
    
    \caption{Left Panel: Image from HST. Middle: Model generated with GALFIT. Right: Residual image after model is subtracted from the data. The far right image shows the 1D isophotal fitting, with the residuals shown in the bottom.}
    \label{fig:galaxy-fits-1}
\end{figure*}

NSAID 67333: This galaxy is best modeled with a single bulge component and a PSF. Interestingly, this galaxy has no emission lines in its spectra and the central object has colors and luminosities consistent with a NSC. 

NSAID 124477: This galaxy is fit with a disk, bulge and a nuclear component, rather than a PSF. There is a clear spiral arm structure apparent in the residuals, we elected not to model the arms. The central point source has properties that are consistent with what we would expect from a NSC. 

NSAID 88260: This galaxy has irregular morphology and was modeled with {two sersic components with fourier modes enabled. We allowed the disk component sersic index to vary in this galaxy due to its irregular structure. The F606W image could only be successfully fit with a single sersic component.} There are tidal features in the image indicating that it may be undergoing some sort of interaction. The point source in this galaxy does not have colors consistent with being an NSC. 

NSAID 57867: This is the lowest mass galaxy in the sample and is modeled by a single faint disk component. There is very little visible structure in this galaxy. The brightest source in this galaxy has colors consistent with a very young star cluster, but it is more massive than what we would expect for a galaxy of this size. 

\begin{figure*}
    \centering
    \includegraphics[trim={8cm 0 8cm 0},clip,width=0.7\textwidth]{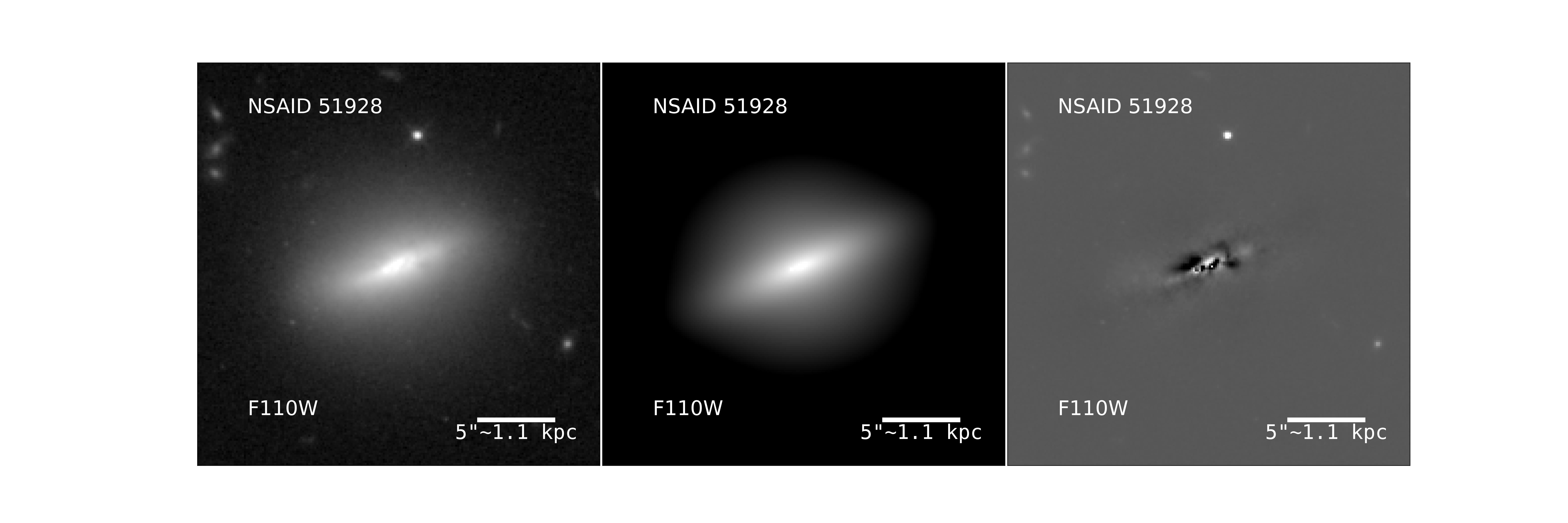}
    \includegraphics[width=0.25\textwidth]{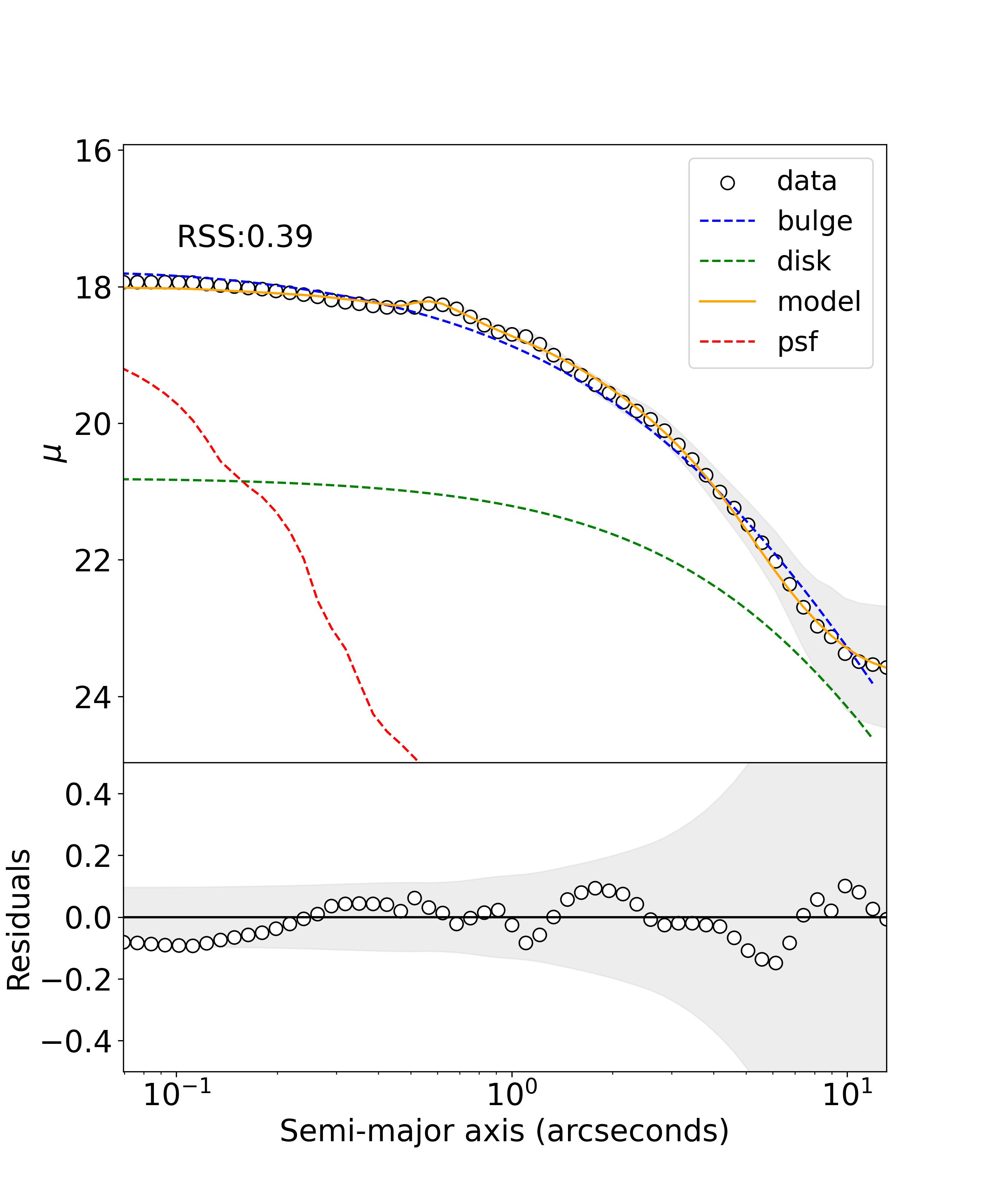} \\

    \includegraphics[trim={8cm 0 8cm 0},clip,width=0.7\textwidth]{124554110_2Dplot.jpeg}
    \includegraphics[width=0.25\textwidth]{124554110_1Dplot.jpeg}\\

    \includegraphics[trim={8cm 0 8cm 0},clip,width=0.7\textwidth]{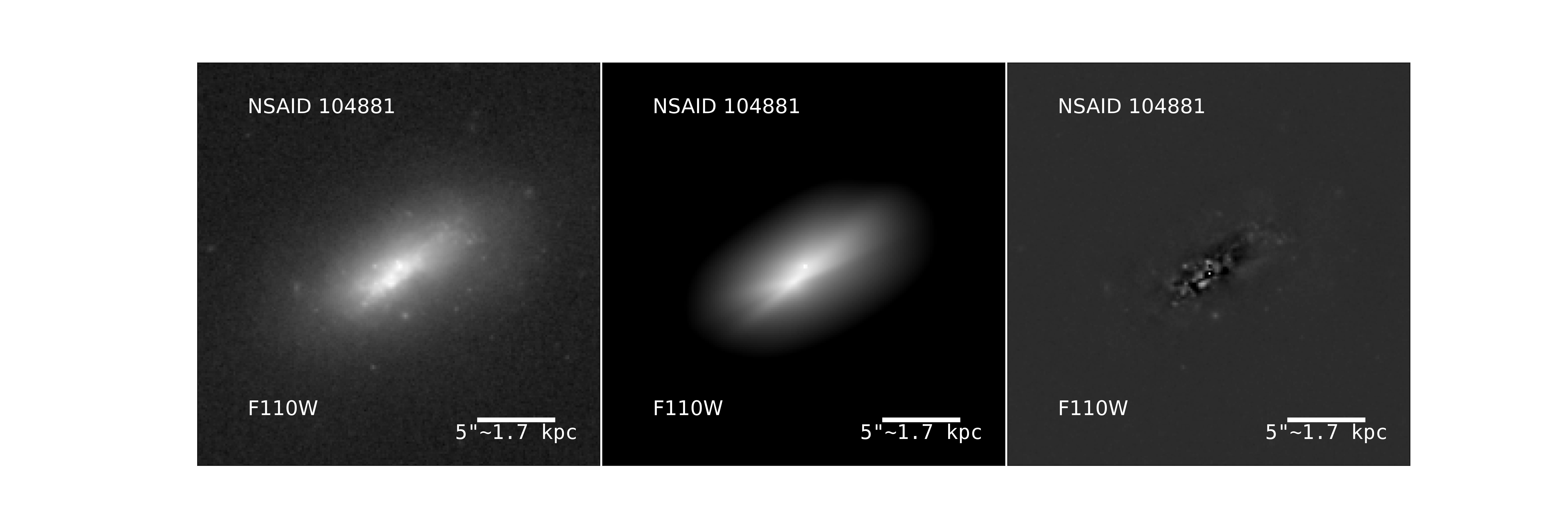}
    \includegraphics[width=0.25\textwidth]{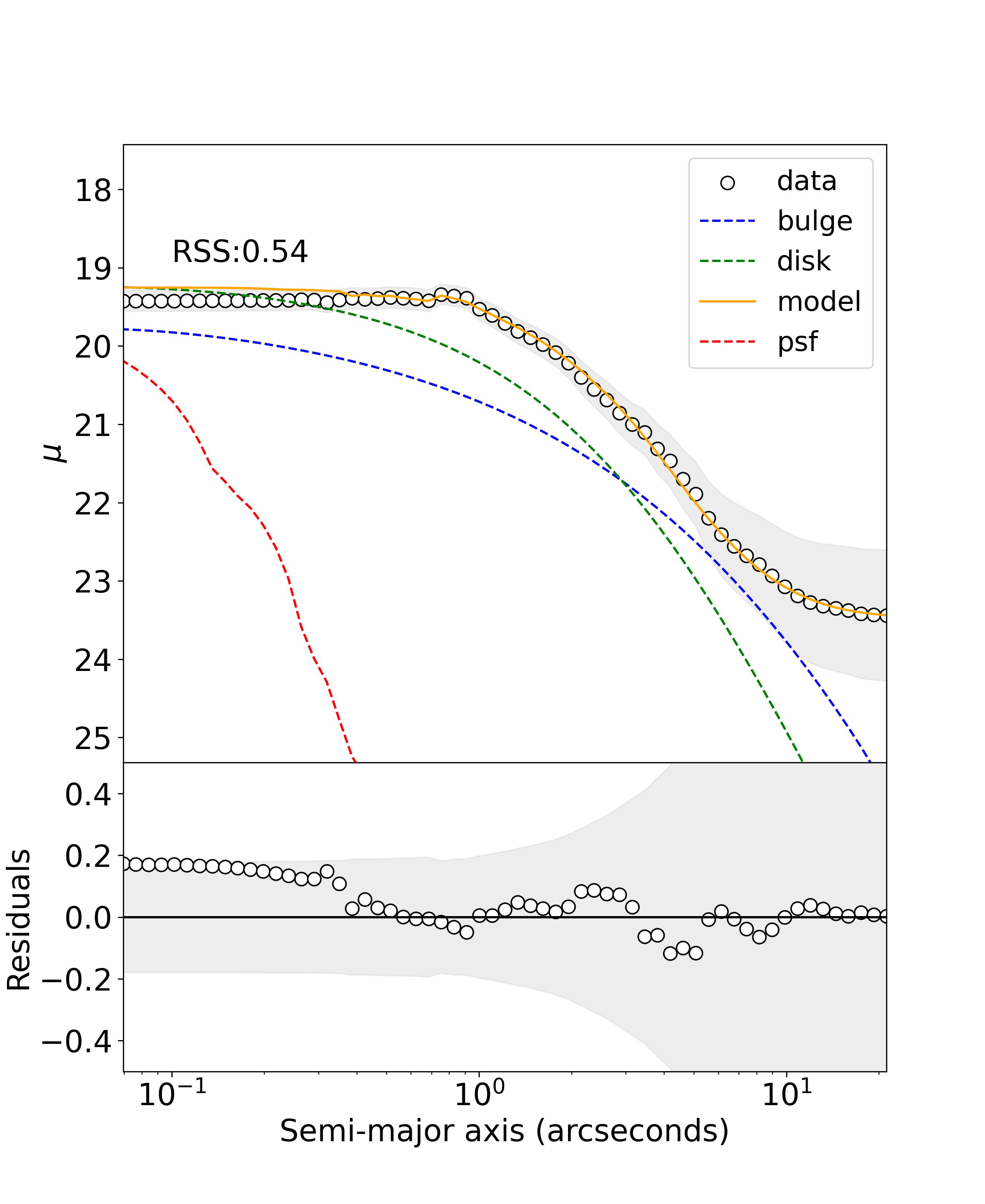}\\

    \includegraphics[trim={8cm 0 8cm 0},clip,width=0.7\textwidth]{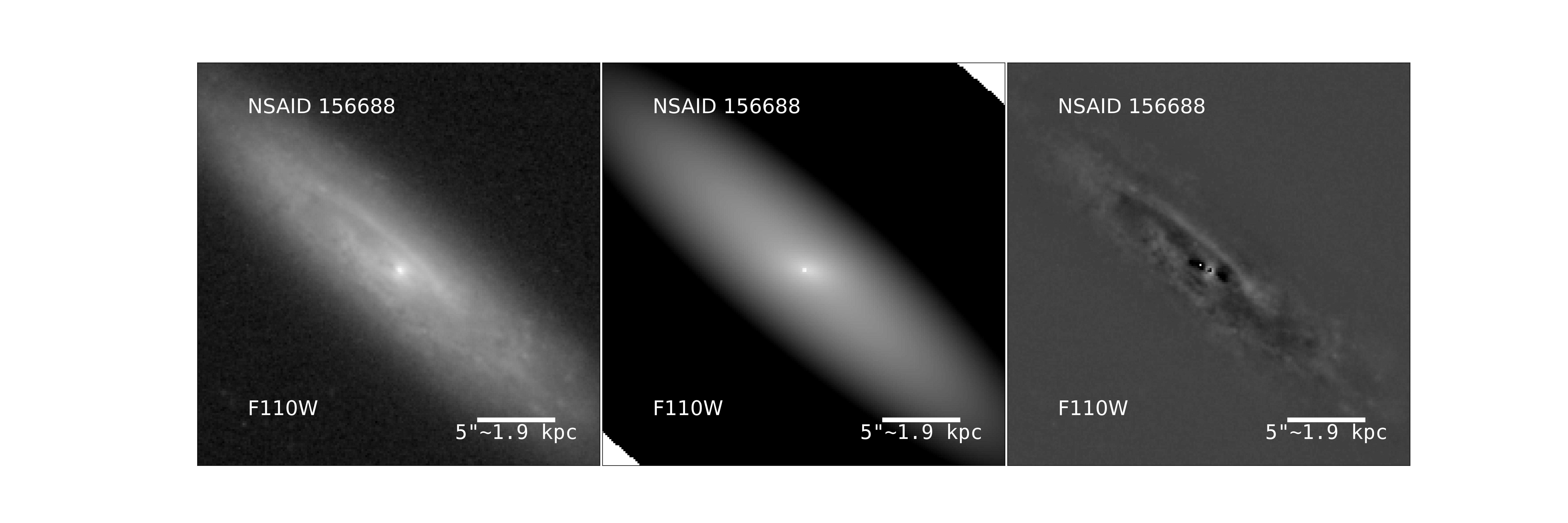}
    \includegraphics[width=0.25\textwidth]{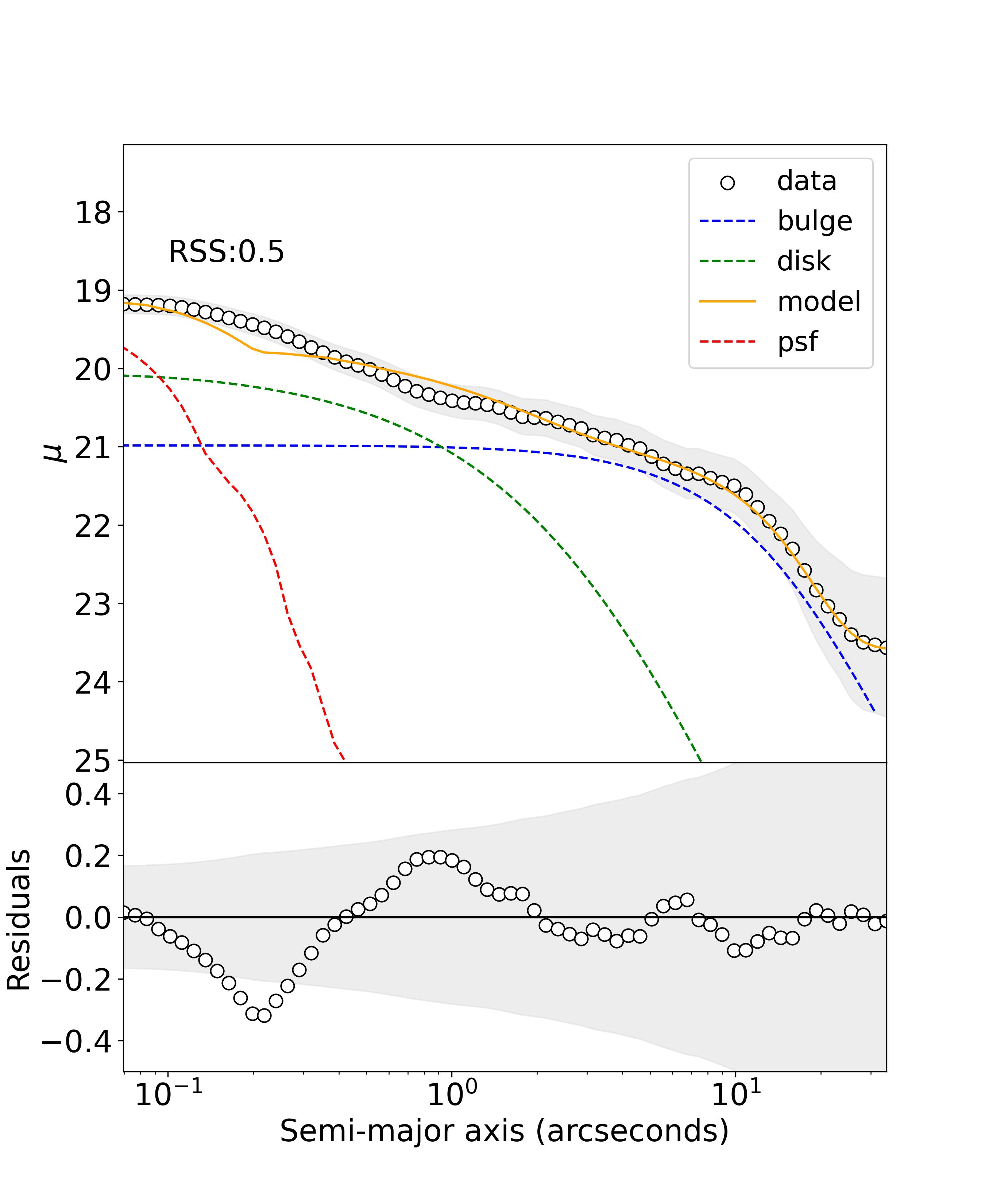}\\

    \caption{Left Panel: Image from HST. Middle: Model generated with GALFIT. Right: Residual image after model is subtracted from the data.The far right image shows the 1D isophotal fitting, with the residuals shown in the bottom.}
    \label{fig:galaxy-fits-2}
\end{figure*}

NSAID 51928: This galaxy is heavily impacted by dust and is best modeled by {two components} and PSF. {The PSF is off nuclear and  the model reflects this by mimicking the bump seen in the data at approximately 1". Because this galaxy has large amounts of dust, the two component model cannot be reproduced in the F606W image. The F606W image is modeled with a single S\'ersic component and a PSF. }This point source does not have properties consistent with being a NSC. 

NSAID 124554: This galaxy is best modeled by a disk, bulge, a nuclear component and PSF. The point source in this galaxy is slightly redder than what we would expect for a NSC, but does not have luminosities that exceed that of a star cluster. 

NSAID 104881: This galaxy is irregular and is best modeled with a disk, bulge and PSF. {To better capture the irregularness of the galaxy we enabled fourier modes in GALFIT}. This galaxy has significant X-ray emission consistent with AGN activity, and the point source has colors not consistent with a NSC. 

NSAID 156688: This galaxy could not be properly modeled in any filter besides F110W due to the extreme dust extinction in this galaxy. Only a faint disk could be successfully fit in the F606W filter, and thus, we excluded this galaxy from the majority of the analysis.

\end{document}